\documentclass[fleqn,10pt]{wlscirep}
\usepackage[utf8]{inputenc}
\usepackage[T1]{fontenc}
\usepackage{nameref}
\usepackage{hyperref}
\renewcommand\thesection{\hspace{-2.3mm}}
\renewcommand\thesubsection{\hspace{-1.8mm}}
\renewcommand\thesubsubsection{\hspace{-1.6mm}}

\title{Circulation of a digital community currency}

\author[1,*]{Carolina E.S. Mattsson}
\author[2,3]{Teodoro Criscione}
\author[1]{Frank W. Takes}
\affil[1]{Leiden Institute of Advanced Computer Science, Leiden University, 2333 CA Leiden, NL}
\affil[2]{Department of Network and Data Science, Central European University, A-1100 Wien, AT}
\affil[3]{Freiburg Institute for Basic Income Studies, University of Freiburg, 79098 Freiburg, DE}

\affil[*]{e.s.c.mattsson@liacs.leidenuniv.nl\\mattsson.c@northeastern.edu}


\begin{abstract}

Circulation is the characteristic feature of successful currency systems, from community currencies to cryptocurrencies to national currencies. In this paper, we propose a network analysis approach especially suited for studying circulation given a system's digital transaction records. Sarafu is a digital community currency that was active in Kenya over a period that saw considerable economic disruption due to the COVID-19 pandemic. We represent its circulation as a network of monetary flow among the 40,000 Sarafu users. Network flow analysis reveals that circulation was highly modular, geographically localized, and occurring among users with diverse livelihoods. Across localized sub-populations, network cycle analysis supports the intuitive notion that circulation requires cycles. Moreover, the sub-networks underlying circulation are consistently degree disassortative and we find evidence of preferential attachment. Community-based institutions often take on the role of local hubs, and network centrality measures confirm the importance of early adopters and of women's participation. This work demonstrates that networks of monetary flow enable the study of circulation within currency systems at a striking level of detail, and our findings can be used to inform the development of community currencies in marginalized areas.

\end{abstract}

\begin{document}

\flushbottom
\maketitle
%
%
\thispagestyle{empty}

\section{Introduction}

The circulation of money is generally studied in an abstract sense, for example as the extent to which monetary policy, productivity improvements, supply disruptions, or other shocks affect aggregate indicators of economic output~\cite{nakamura_identification_2018,mcnerney_how_2022,carvalho_supply_2016,acemoglu_network_2012}. Detailed observation has long been impractical for lack of empirical data. However, modern payment infrastructure is increasingly digital~\cite{adrian_rise_2019}, and the circulation of money is leaving real-time records on the servers of financial institutions worldwide. These transaction records offer especially high granularity data in time and in space, and open up the possibility of fined-grained data-driven studies of financial ecosystems~\cite{frankova_transaction_2014,alessandretti_anticipating_2018,aladangady_transactions_2019,bouchaud_trades_2018,mattsson_trajectories_2021,bardoscia_physics_2021,carvalho_tracking_2021}. In this paper we consider the circulation of money as observed in transaction records, using \emph{networks of monetary flow} to represent patterns of circulation over a period of time. Our findings show that techniques in network science --- flow-based community detection, measures of cyclic structure, network mixing patterns, and walk-based centrality metrics --- together capture key aspects of circulation within a real-world currency system when applied to such networks. We demonstrate that important practical and theoretical questions around the circulation of money can be studied using networks of monetary flow. 

The main focus of this paper is on complementary currencies whose modern implementations produce comprehensive digital records, as these are cases where transaction records are available for an entire currency system. Complementary currencies circulate in parallel to national currencies in that tokens are \textit{not} legal tender, nor necessarily exchangeable for legal tender~\cite{stodder_complementary_2009,lietaer_complementary_2004,ussher_complementary_2021}; they are used under mutual agreements that come in many forms, from local community currencies~\cite{muralt_woergl_1934,kichiji_network_2008,frankova_transaction_2014} to global cryptocurrencies~\cite{nakamoto_bitcoin_2008,kondor_rich_2014,elbahrawy_evolutionary_2017}. Sardex, for example, is a digital complementary currency used among businesses in Sardinia. Digital records of transactions in Sardex have been studied to show that cycle motifs are related to performance and stability of the currency system~\cite{iosifidis_cyclic_2018,fleischman_balancing_2020}. The full transaction histories of Bitcoin and other cryptocurrencies can be reconstructed from public ledgers~\cite{ober_structure_2013,meiklejohn_fistful_2016,zhang_heuristic-based_2020}. Bitcoin transactions reveal a currency system that supports substantial trade outside centralized marketplaces, but where inequality has been increasing over time~\cite{nadini_emergence_2022,kondor_rich_2014}. Sarafu, the currency considered in this work, is a ``Community Inclusion Currency'' (CIC) that incorporates elements of both community currencies and cryptocurrencies~\cite{mattsson_sarafu_2022}. 

Community currencies are typically created with the aim to support local economic activity by facilitating local circulation~\cite{muralt_woergl_1934,kichiji_network_2008,ruddick_eco-pesa_2011}. Moreover, complementary currencies are counter-cyclical in that they tend to see higher circulation during periods of economic disruption~\cite{stodder_macro-stability_2016,zeller_economic_2020}. We study Sarafu, a digital community currency with a presence in several local areas of Kenya, from January 2020 to June 2021, a period that saw considerable economic disruption due to the COVID-19 pandemic. Previously published data~\cite{ruddick_sarafu_2021} and a data descriptor~\cite{mattsson_sarafu_2022} detail the transaction activity that occurred over Sarafu over this period. We detail transaction volumes in Sarafu over time and then study the circulation of Sarafu resulting from this activity as a network of monetary flow.

The Sarafu flow network encompasses 293.7 million Sarafu in transaction volume among around 40,000 users. This weighted, directed, time-aggregated network representation captures the patterns of circulation in intricate detail, allowing us to study what shapes the Sarafu currency system as a whole. We apply carefully selected network analysis techniques to the Sarafu flow network to answer three questions of particular relevance to research about community currencies: 

\textit{Among whom is Sarafu circulating?} Anonymized information on account holders allows us to label each node with a geographic area, livelihood category, registration date, and reported gender. We investigate the composition of sub-populations within which Sarafu was circulating, as identified using a so-called community detection method developed especially for flow networks. Specifically, the map equation framework and the associated Infomap algorithm~\cite{rosvall_maps_2008,ding_community_2014} group nodes into sub-populations whose sub-networks capture as much of the transaction volume as possible.

\textit{What network structures support the circulation of Sarafu?} Degree disassortativity has been noted in a variety of economic networks~\cite{fujiwara_large-scale_2010,kondor_rich_2014,mattsson_functional_2021,campajola_microvelocity_2022} in that high-degree nodes generally transact with low-degree nodes. It has also been noted that network cycles may be key to the `health' of currency systems and of individual accounts\cite{iosifidis_cyclic_2018}. Indeed, detecting cycles and brokering `missing' financial connections is seen by private actors as a promising credit clearing and risk management service~\cite{fleischman_balancing_2020,fleischman_liquidity-saving_2020}. In a similar vein, Ussher et al.\cite{ussher_complementary_2021} argue that community currencies compare favorably to cash assistance as an economic development intervention because they help establish economic connections that keep money local. We study the network structure underlying the circulation of Sarafu within the aforementioned sub-populations using network assortativity measures and relative cycle counts. 

\textit{What characterizes the most prominent Sarafu users?} We would like to understand patterns in who holds Sarafu accounts that are especially prominent, or perhaps even systematically important. Prominent users are identified by means of a network centrality measure that is directly related to the circulation of Sarafu, as captured by a network of monetary flow. Specifically, weighted PageRank~\cite{page_pagerank_1999} computes a metric that corresponds to the share of funds a given account would control, at any given time, if the observed dynamics were to continue indefinitely. We calibrate this measure against empirical account balances and use it to investigate the account features most associated with prominent users.

The Sarafu user base grew substantially over the observation period and transaction volumes rose dramatically as the COVID-19 pandemic disrupted regular economic activities. However, our results indicate that circulation remained modular and geographically localized. Circulation occurred primarily within particular sub-populations consisting of co-located users with diverse livelihoods. This has concrete implications for humanitarian policy in marginalized areas, in that it implies that community currencies can help support specific areas during periods of economic stress and that such interventions are more likely to succeed in areas with a diverse mix of economic activities already present. Community currencies also support local economic development over longer periods of time~\cite{stodder_complementary_2009,iosifidis_cyclic_2018,ussher_complementary_2021}. For such initiatives, it is useful to know that community-based financial institutions, and, in a few cases, faith leaders, are especially prominent among Sarafu users. These ``hubs'' play a key structural role especially at a local level, in that the sub-networks underlying local circulation are consistently degree-disassortative. Moreover, an elevated presence of short cycles supports the intuitive notion that circulation requires cycles.

The findings and insights presented in this paper provide a fine-grained understanding of the circulation of Sarafu over a highly dynamic period that includes the arrival of the COVID-19 pandemic to Kenya. This investigation demonstrates that networks of monetary flow can capture key features of circulation within currency systems. Moreover, flow-, walk-, and cycle- based network measures and algorithms produce interpretable analyses that allow for characterising the system. Noteworthy is that the approach presented in this paper can be applied to study any currency system where digital transaction records are available. Indeed,there may be be important regularities underlying the circulation of money in such systems, and these would be well worth exploring further.

The remainder of this paper is organized as follows. The \nameref{sec:data} section briefly describes the Sarafu system over this especially tumultuous period. The \nameref{sec:results} section presents our findings on patterns of circulation, prominent users, and the network structure underlying circulation. We synthesize these contributions and discuss the implications of our findings in the \nameref{sec:discussion} section. Finally, the \nameref{sec:methods} section details the data preparation, network analysis measures, and statistical methods used in this study and provides references to facilitate data, code, and software availability.

\section{Data}\label{sec:data}

Digital administrative records of the Sarafu CIC from January 2020 to June 2021 have been published by Grassroots Economics, a non-profit foundation based in Kenya that operates Sarafu and leads related economic development projects in marginalized and food-insecure areas of the country. The published dataset\cite{ruddick_sarafu_2021} has previously been described in raw form\cite{mattsson_sarafu_2022} and used in a case study introducing CICs as a modality for humanitarian aid~\cite{ussher_complementary_2021}. During the observation period, Sarafu was available throughout Kenya. Mimicking the well-developed mobile payment infrastructure of the national currency~\cite{mbogo_impact_2010,stuart_cash_2011,mbiti_mobile_2011,suri_mobile_2017,koblanck_digital_2018,baah_state_2021}, each Sarafu account was tied to a Kenyan mobile number and accessible over a mobile interface. An account could be created with an activation code sent to a particular mobile number, then used and managed via a series of simple menus. The resulting digital records became a dataset that includes hundreds of thousands of Sarafu transactions and anonymized account information for the tens of thousands of users.

\begin{figure}[!t]
\centering
\includegraphics[width=0.9\textwidth]{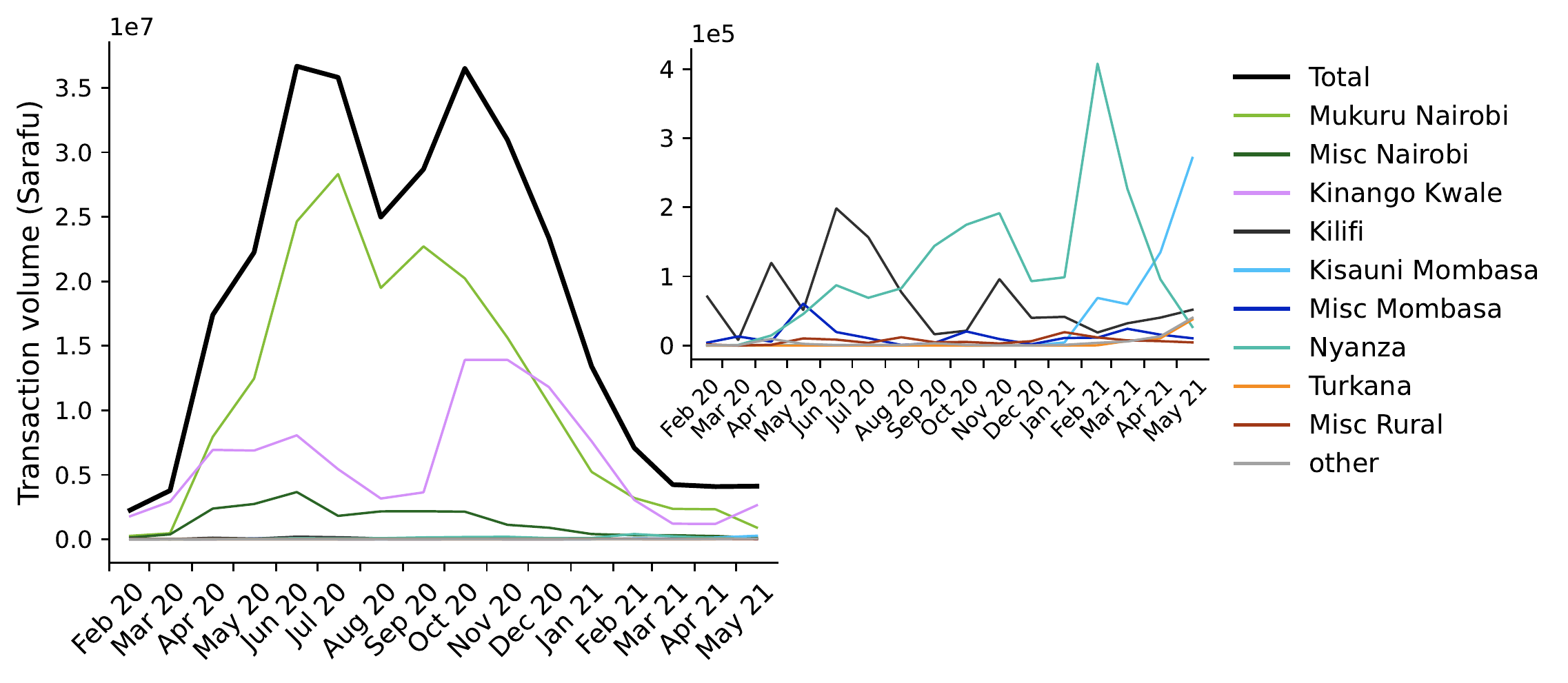}
\caption{Monthly transaction volumes in total, and in each geographic area (shown at two different scales).}
\label{fig:sarafu}
\end{figure}

January 2020 saw the consolidation of several precursor currencies onto a single platform: Sarafu. In prior years, several digital community currencies had been launched by Grassroots Economics in different areas\cite{marion_voucher_2018,mattsson_sarafu_2022}. These were designated to operate as separate systems over a decentralized infrastructure, interacting very little. Flow networks from earlier periods would have been composed of a dozen or so nearly-disconnected components, each corresponding to a different precursor currency serving a different local community. Shortly following the start of the observation period, the consolidated system expanded dramatically as the COVID-19 pandemic arrived in Kenya. Sarafu grew from 8,354 registered accounts in January 2020 to almost 55,000 in June 2021. Figure~\ref{fig:sarafu} shows the transaction volumes for each of the complete months over the observation period. Beginning in April 2020 and continuing through the second wave of COVID-19 in Kenya, Sarafu saw transaction volumes almost ten times higher than in February 2020. This dramatic expansion occurred primarily in particular regions, described below, and we see a return towards the baseline in these areas by the end of the observation period. The overall pattern is perhaps best explained by the counter-cyclical nature of complementary currencies, which are known to see spikes in usage levels during periods of economic disruption~\cite{stodder_complementary_2009,stodder_macro-stability_2016,zeller_economic_2020}. 

The figures in this work employ a consistent color scheme for geographic area. Purple corresponds to \emph{Kinango Kwale}, a rural area where GE has had a substantial presence in several local communities for many years; this area saw much growth during the COVID-19 pandemic due largely to word of mouth. Light green is \emph{Mukuru Nairobi}, an urban area that was the site of a targeted introduction beginning in March 2020. For details we refer to the \nameref{sec:methods:data} section in \nameref{sec:methods}. Accounts located elsewhere in \emph{Nairobi} are shown in dark green. In light blue are accounts in \emph{Kisauni Mombasa}, the site of a second introduction beginning in early 2021. Accounts located elsewhere in \emph{Mombasa} are shown in dark blue. \emph{Kilifi}, in dark grey, is the county where GE is headquartered. Users with an unknown location (the largest category within \emph{other}), are in light grey. In Figure~\ref{fig:sarafu}, \emph{other} closely tracks the remote rural county of \emph{Turkana}, in orange. Teal and red correspond to locations in \emph{Nyanza} county or elsewhere in rural Kenya, respectively.

\section{Results} \label{sec:results}

\begin{figure}[ht!]
    \centering
  \includegraphics[width=6in]{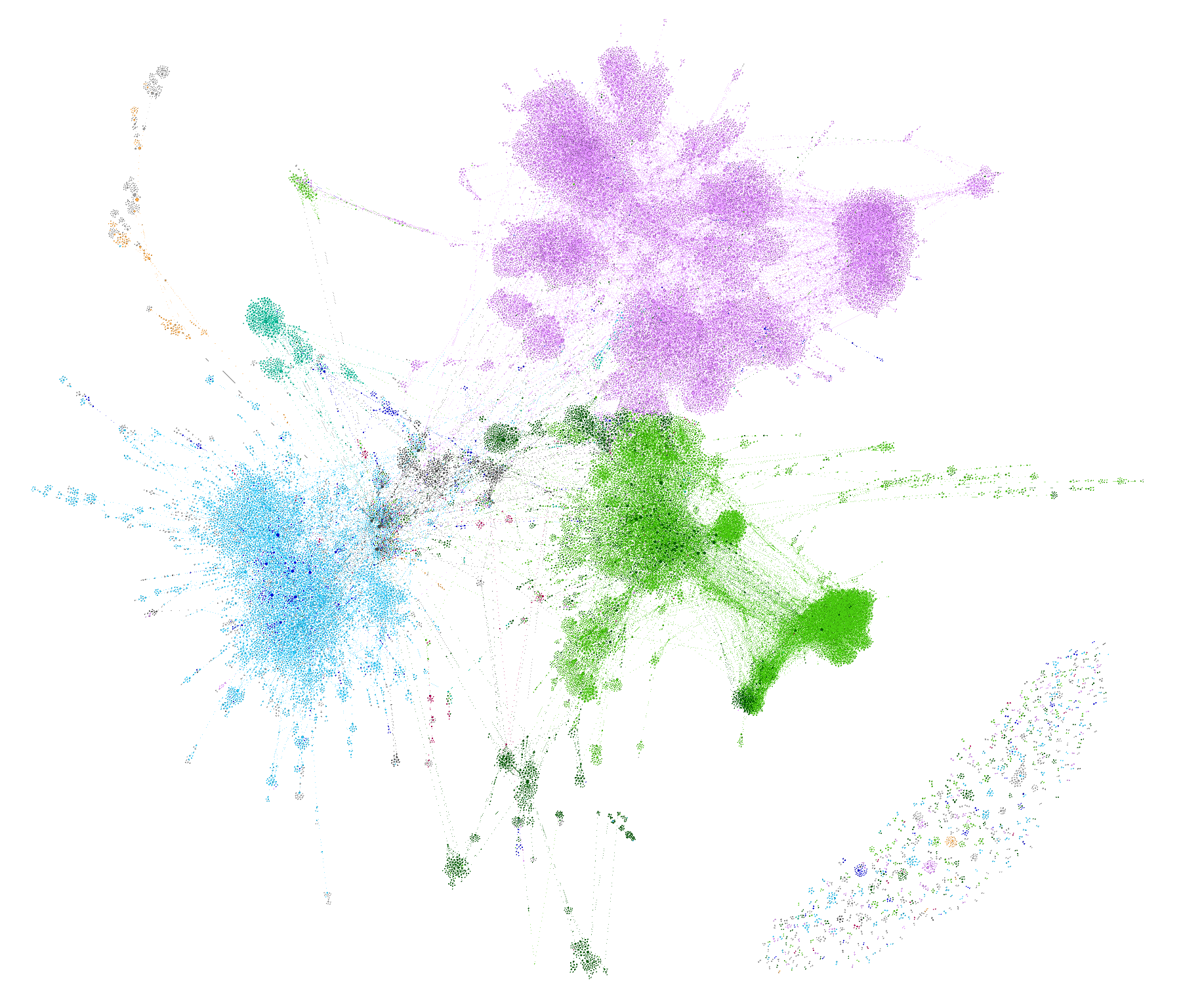}
  \caption{Visualization of the Sarafu flow network. Nodes are colored by the geographic area of the location reported for the account (see Figure~\ref{fig:sarafu} for legend), and node size is proportional to the value of unweighted PageRank as computed for that node.}
  \label{fig:network}
\end{figure}

The Sarafu system supported over 400,000 transactions among more than 40,000 user accounts between January 2020 and June 2021. This resulted in the circulation of 293.7 million Sarafu, visualized in Figure~\ref{fig:network} as a network of monetary flow. The \emph{nodes} are registered accounts, for which we know attributes such as the geographic area, livelihood category, and reported gender of the account holder. An \emph{edge} from one account to another indicates that at least one transaction occurred across that link. The \emph{edge weight} corresponds to the observed monetary flow along an edge, i.e., the total sum of transaction amounts across that link. The Sarafu flow network is a \emph{weighted, directed, time-aggregated network representation} of the total circulation over the observation period, excluding system-run accounts. For details on the construction of the network, we refer to the \nameref{sec:methods:data} section of \nameref{sec:methods}. The network visualization employs the same colors for geographic area as does Figure~\ref{fig:sarafu}, revealing patterns suggestive of modular and geographically localized circulation.

In the remainder of this section, we share findings resulting from network analysis of the Sarafu flow network. The \nameref{sec:results:flow} section considers sub-populations within which Sarafu was circulating, and their composition along lines of geographic area and livelihood category. In the \nameref{sec:results:structure} section, we consider the network structure that supports this circulation, including analyses of cyclic structure and network mixing patterns. The \nameref{sec:results:hubs} section compares relevant network centrality measures and describes the most prominent users of Sarafu. 

\subsection{Modular circulation} \label{sec:results:flow}

To more precisely understand the patterns of circulation present in the Sarafu flow network, we apply an especially suitable community detection method. The map equation\cite{rosvall_maps_2008} is defined in terms of flow networks and the associated Infomap algorithm\cite{ding_community_2014} groups nodes into hierarchical \emph{modules}. Specifically, Infomap assigns nodes to modules (and sub-modules) within which a ``random walker'' on the network would stay for relatively long periods of time. In our case, the weights on the edges of the Sarafu flow network reflect real, observed flows of Sarafu and so the Infomap algorithm will seek to discover modules that contain especially much transaction volume. This identifies sub-populations within which Sarafu tended to \emph{circulate}. For details, we refer to the \nameref{sec:methods:analysis:circ} section in \nameref{sec:methods}.

The Infomap algorithm recovers a hierarchical, nested, modular structure to the Sarafu flow network. The hierarchical structure consists of top-level modules, sub-modules and sub-sub-modules at respectively the first, second and third level of the community hierarchy. Circulation of the Sarafu community currency was highly modular in that activity occurred almost exclusively within distinct sub-populations. At the first hierarchical level, 99.7\% of the total transaction volume was contained within the five largest so-called \emph{top-level modules}. Moreover, there are 37 \emph{sub-modules} composed of 100 or more accounts and these contained 96.5\% of the total transaction volume. Only a small share of the overall circulation took place between the sub-populations defined at the second hierarchical level, and circulation within these sub-populations itself had a nested, modular structure. Indeed, the 455 \emph{sub-sub-modules} composed of 10 or more accounts capture 80\% of the total transaction volume. Altogether, these findings suggest that the circulation of Sarafu was extremely modular over the observed period.

\subsubsection*{Geographic localization} \label{sec:results:geog}

\begin{figure}[!t]
    \centering
  \includegraphics[width=0.45\textwidth]{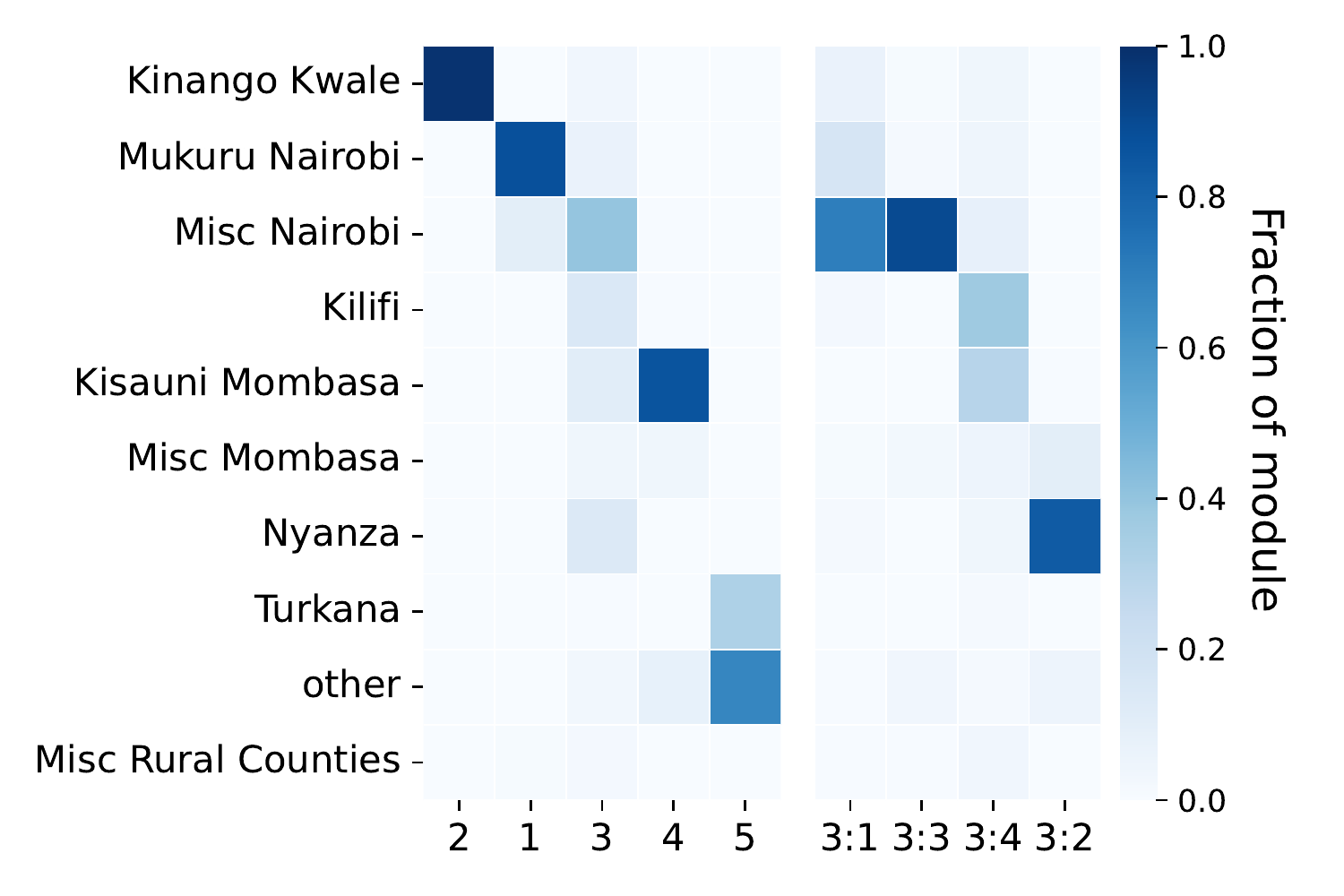}
  \caption{Geographic composition of the five largest top-level modules and relevant numbered sub-modules.}
  \label{fig:matrix}
\end{figure}

We investigate the extent to which the distinct sub-populations discovered above localize in particular geographic areas. Figure~\ref{fig:matrix} shows the geographic composition of the top-level modules---four of the five map directly onto one of the main areas labeled in the data: \textit{Kinango Kwale}, \textit{Mukuru Nairobi}, \textit{Kisauni Mombasa}, or \textit{Turkana}. Only one of the modules has substantial membership from several regions; its sub-modules are, however, also geographically delineated. This top-level module combines several less prominent localities, including in \emph{Kilifi}, in \emph{Nyanza}, and in two localities elsewhere in \emph{Nairobi}. We conclude that the circulation of Sarafu was geographically localized over the observed period.

The top-level modules are amalgamations of circulation within sub-modules, which appear to correspond to geographic areas more granular than those labeled in the data. Indeed, raw reported locations were often quite precise and were converted to broader area labels in the anonymization that occurred prior to the publication of the data~\cite{mattsson_sarafu_2022}. Several of the sub-modules highlighted in Figure~\ref{fig:matrix} coincide with areas where early, physical community currencies were operating in the years before Sarafu became all-digital~\cite{marion_voucher_2018}. Within \emph{Kinango Kwale}, moreover, the sub-modules likely correspond to individual rural villages or clusters of villages~\cite{ussher_complementary_2021}. Thus, circulation was geographically local, predominantly. We will further consider the sub-populations delineated by the Infomap sub-modules in subsequent analyses.

\subsubsection{Diversity of economic activities} 
\label{sec:results:uses}

Now that we understand the modular structure and geographic localization of circulation, we consider the composition of the localized sub-populations with respect to economic activity. This is of particular interest to practitioners as it helps illustrate \emph{among whom} Sarafu was circulating. There are 14 categories of economic activities into which user-reported livelihoods were grouped, the most common of which are \emph{labour} in urban areas and \emph{farming} in rural areas. Many other users (in both urban and rural areas) report selling \emph{food}, running a \emph{shop}, or providing \emph{transport}. 

\begin{figure}[ht]
  \includegraphics[width=0.8\textwidth]{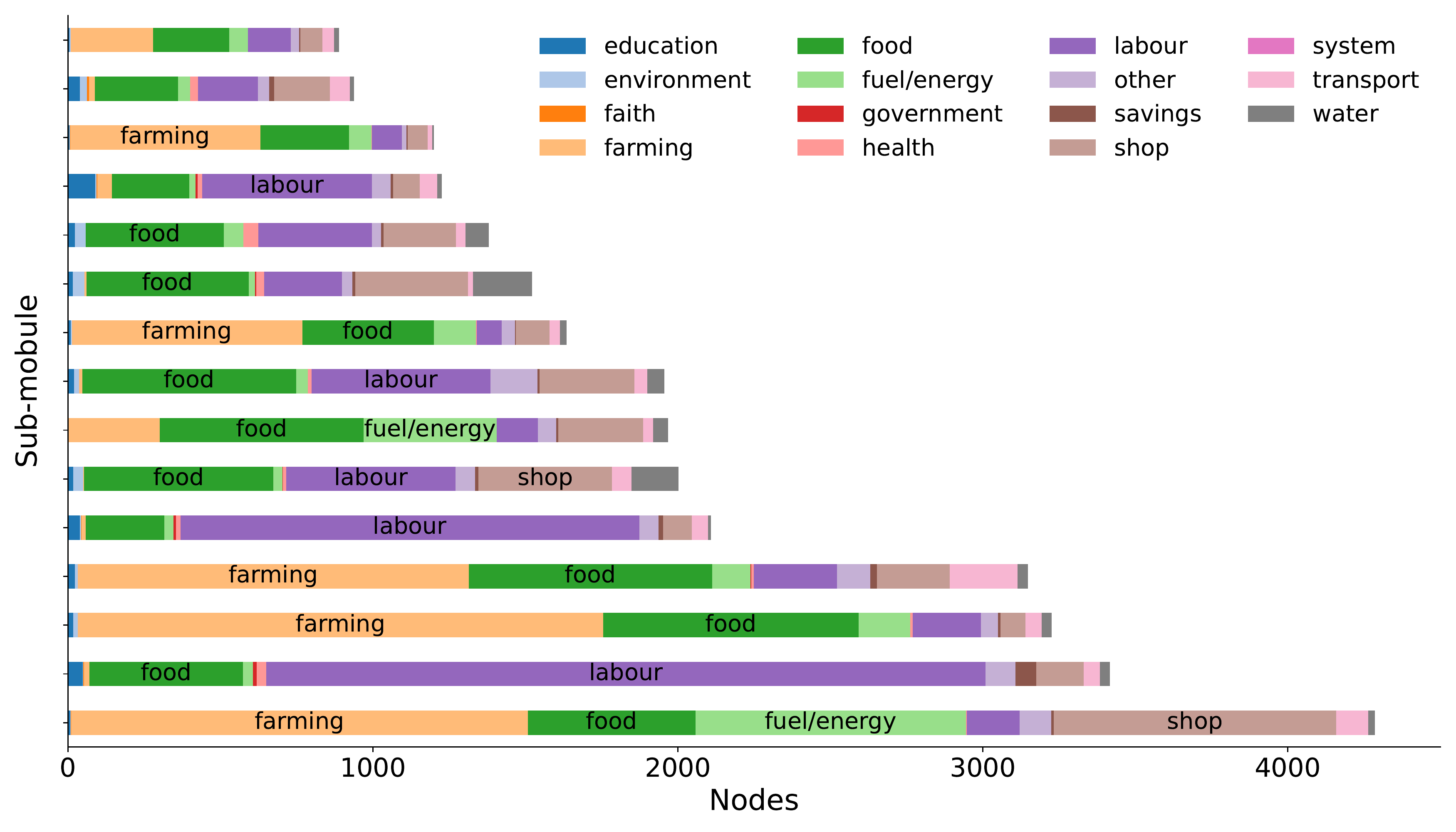}
  \caption{Composition of discovered sub-modules (bars) in terms of user livelihoods (colors, as shown in legend).}
  \label{fig:uses}
\end{figure}

Most notably, we see a mix of the different economic activities within the largest second-level sub-populations. Figure~\ref{fig:uses} illustrates the livelihood category given for each account in the 15 largest sub-modules identified by the Infomap algorithm. The largest sub-populations include several thousand accounts, and 13 of the 37 we consider have more than 1000 users. Notice that the livelihood mix consistently diverse. To also give a sense of how this diversity is experienced from within sub-populations, we compute and report the view from the average user. The average user participates in a sub-module with around 2000 other users, and of these others, 66\% report a category of work that is different from what they themselves report. Little diversity is lost as we consider even finer scales. The average user appears in a sub-sub-module with around 250 other users, 59\% of whom do not share their same livelihood category. We conclude that the circulation of Sarafu involves a diversity of economic activities, even at the scale of a single village.

We also see that the composition of the sub-populations using Sarafu is substantively different in urban and rural areas. In Figure~\ref{fig:uses}, the sub-modules where \emph{farming} or \emph{fuel/energy} are prominent are rural and composed of users reporting a location within \textit{Kinango Kwale}, almost exclusively. Those where \emph{labour} is prominent correspond to sub-populations localized primarily in urban or peri-urban areas including \textit{Mukuru Nairobi}, \textit{Kisauni Mombasa}, and \textit{Kilifi}. The geographic aspect of circulation is further refined by means of the type of geographical area.

\subsection{Underlying network structure}  \label{sec:results:structure}

In this section, we consider the network structure underlying the circulation of Sarafu. Each of the sub-modules considered above in the~\nameref{sec:results:flow} section is associated with a sub-population of 100 or more accounts that defines a sub-network of 100 or more nodes. An (unweighted) edge from one account to another indicates that at least one transaction occurred across that edge. Node degree corresponds to an accounts' number of unique transaction partners, incoming and outgoing, in their same sub-population. In the \nameref{sec:results:structure:cycles} section we count the cycles present in the sub-networks, relating the presence of cycles to the notion of circulation and the sustainable operation of complementary currency systems. Next, the \nameref{sec:results:structure:mixing} section quantifies network mixing patterns, relating degree disassortativity to the structural importance of local ``hubs'' in the sub-networks.

\subsubsection{Cyclic structure} \label{sec:results:structure:cycles}

Network cycles may be key to understanding the conditions under which an area is, or overtime becomes, able to sustain local circulation\cite{iosifidis_cyclic_2018,fleischman_balancing_2020,ussher_complementary_2021}. We explore the presence of cycles in the Sarafu sub-networks as compared to the expectation from a null model. We use two standard null models: Erdős-Rényi (ER) networks and random degree-preserving (RD) networks. ER networks have the same number of nodes and edges as the empirical network, but are wired randomly. Our RD networks preserve the indegree and outdegree sequences in expectation. For details we refer to the \nameref{sec:methods:analysis:cycles} section of \nameref{sec:methods}.

\begin{figure}[htb]
\centering
\includegraphics[width=0.7\textwidth]{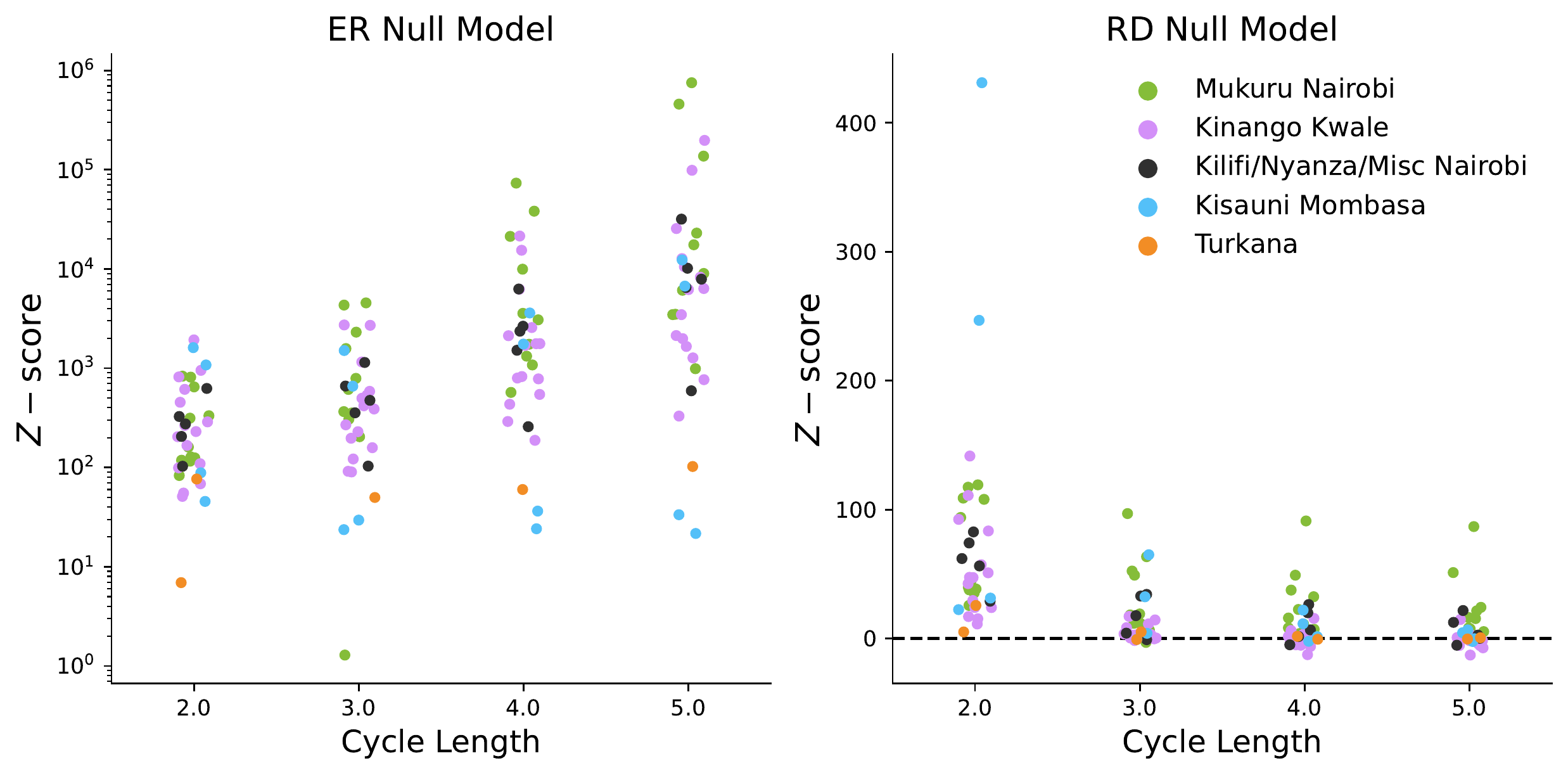}
\caption{Relative cycle counts for each sub-network, at different cycle lengths. Points correspond to sub-modules, and are colored based on the dominant geographic area of users placed in the top-level module to which it belongs. Five observations are omitted from the ER plot, as the logarithmic scale cannot represent small negative $z$-scores. These are confined to two sub-modules, where one or no cycles occur at higher cycle lengths.}
\label{fig:kcycle_submod}
\end{figure}

Relative to the ER null model, nearly all of the sub-networks have many more cycles than expected at cycle lengths $2$, $3$, $4$, and $5$ (Figure~\ref{fig:kcycle_submod}, left). Extremely large magnitudes of the $z$-score may indicate that ER networks are not a suitable comparison. Indeed, short cycles are negligible in large Erdős-Rényi networks\cite{bianconi_local_2008}. Figure~\ref{fig:kcycle_submod} (right) shows the $z$-score of the cycle counts computed for each of the Sarafu sub-networks, relative to the RD null model. The values are positive and large for cycles of length $2$ and $3$ for most sub-networks, indicating cycle counts many standard deviations above expectation. Some sub-networks, especially in \textit{Mukuru Nairobi}, also have unexpectedly many cycles of length $4$ and $5$. These findings are in line with prior findings for the Sardex currency in Sardinia, where a gradual increase in short cycles (relative to an RD null model) was observed over two years\cite{iosifidis_cyclic_2018}. Notably, this is the case even though Kenya and Sardinia differ in their level of economic development, Sardex is aimed at businesses whereas Sarafu is aimed at individuals, and pandemic times are certainly different from regular times. Moreover, the currency management practices followed by the two providers are quite distinct\cite{iosifidis_cyclic_2018,mattsson_sarafu_2022}. Short cycles are a prominent network connectivity pattern in the circulation of Sarafu, and perhaps of (community) currencies more generally.

\subsubsection{Structural correlations}
\label{sec:results:structure:mixing}

Degree disassortativity is an expected feature of specifically currency networks~\cite{kondor_rich_2014,campajola_microvelocity_2022} and of economic networks, more broadly~\cite{fujiwara_large-scale_2010,mattsson_functional_2021}. In networks with this property, high-degree nodes generally interact with low-degree nodes, not other high-degree nodes. In avoiding one another, high-degree nodes tend to become ``local hubs'' that are prominent especially with respect to nearby nodes (i.e., their network neighborhood) that are predominantly of low-degree. Recall also that Sarafu sub-modules are diverse with respect to the livelihood reported by accounts (\nameref{sec:results:uses} section). Here we consider these and other structural correlations that help us better understand the circulation of Sarafu. Since the overall influence of account attributes on the Sarafu flow network is limited by the constraints of geography, and may be heterogeneous across sub-populations, we consider degree and attribute assortativities across the Sarafu sub-networks. For details we refer to the \nameref{sec:methods:analysis:mix} section of \nameref{sec:methods}.

Table~\ref{tab:assortativity} reports the average, the median, and the range for each property as computed on the undirected version of each sub-network. We find substantial disassortativity in degree across nearly all sub-networks. As expected, we also find that attribute assortativity is consistently low along the dimension of livelihood category. The consistency of these observations across sub-populations suggests that there may be important regularities in the structural correlations of networks that support the local circulation of community currencies such as Sarafu.

\begin{table}[h]
    \centering
    \caption{Network statistics and feature assortativity across sub-modules with 100 or more nodes.}
    \label{tab:assortativity}
    \begin{tabular}{l|rrr|rrrrr}
            & \multicolumn{3}{|l}{Network Statistics} & \multicolumn{5}{|l}{Assortativity} \\
        & Nodes & Edges & Volume    & Business  & Gender & Registration & Degree & W. Degree \\ \hline
mean    & 1021 &  2636 &  7.67m &  0.032 &  0.146 &  0.154 & -0.215 & -0.066 \\
std	    & 1082 &  3789 & 10.83m &  0.047 &  0.188 &  0.273 &  0.119 &  0.143 \\
min	    &  136 &   170 &  0.01m & -0.104 & -0.081 & -0.323 & -0.448 & -0.428 \\
25\%	&  222 &   544 &  0.59m &  0.003 &  0.017 & -0.072 & -0.265 & -0.168 \\
50\%	&  537 &  1151 &  2.99m &  0.029 &  0.121 &  0.147 & -0.221 & -0.096 \\
75\%	& 1522 &  3513 & 11.05m &  0.058 &  0.208 &  0.322 & -0.152 &  0.012 \\
max	    & 4286 & 20458 & 43.64m &  0.121 &  1.000 &  0.845 &  0.247 &  0.269        
    \end{tabular}
\end{table}

Correlations with respect to \emph{gender} and \emph{registration date} in the structure of the sub-networks can also be substantial, although these effects are not consistent across sub-populations. Again from Table~\ref{tab:assortativity}, attribute assortativity on gender is present in about half of the 37 sub-populations and is substantial in several. This may be related to the activity of community-based savings and investment groups, where women's participation is high\cite{avanzo_relational_2019,rasulova_impact_2017}. Within Sarafu, such groups provide opportunities to transact assortatively on gender. Gender assortativity in payment networks may also reflect, for instance, gendered economic roles in ways that deserve further study. Strong correlations in registration date also appear in several sub-networks, indicating a cohort effect. For example, during targeted introductions as described in the \nameref{sec:methods:data} section, groups of users who share latent economic ties would together adopt Sarafu over a relatively short period of time. Correlations by cohort are likely to appear in any digital payment system where use is voluntary and adoption can be coordinated.

\subsection{Prominent Sarafu users} \label{sec:results:hubs}

Local hubs play a key structural role in the circulation of Sarafu, and it is important to understand who takes on such prominent positions. We ask what features are especially consistent among accounts with high network centrality, now across the entire Sarafu flow network. In the section \nameref{sec:results:hubs:centrality} we consider an account's number of unique transaction partners, transaction volumes, and especially relevant network centrality measures. Next, the \nameref{sec:results:hubs:users} section asks what features of Sarafu accounts are strongly and consistently associated with high network centrality.

\subsubsection{Identifying prominent users} \label{sec:results:hubs:centrality}

\begin{figure}[!t]

  \centering
  \includegraphics[width=0.35\textwidth]{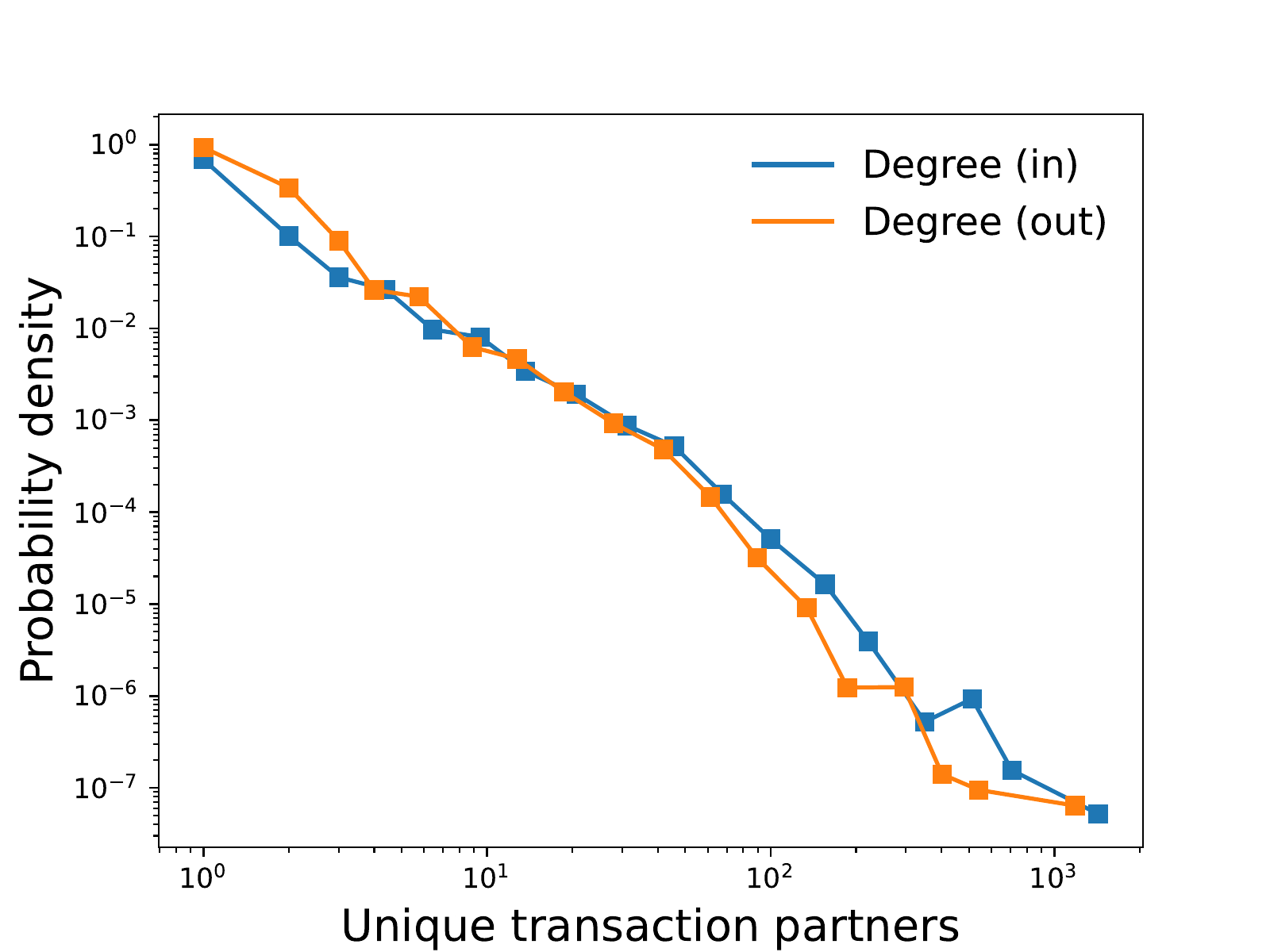}
  \includegraphics[width=0.35\textwidth]{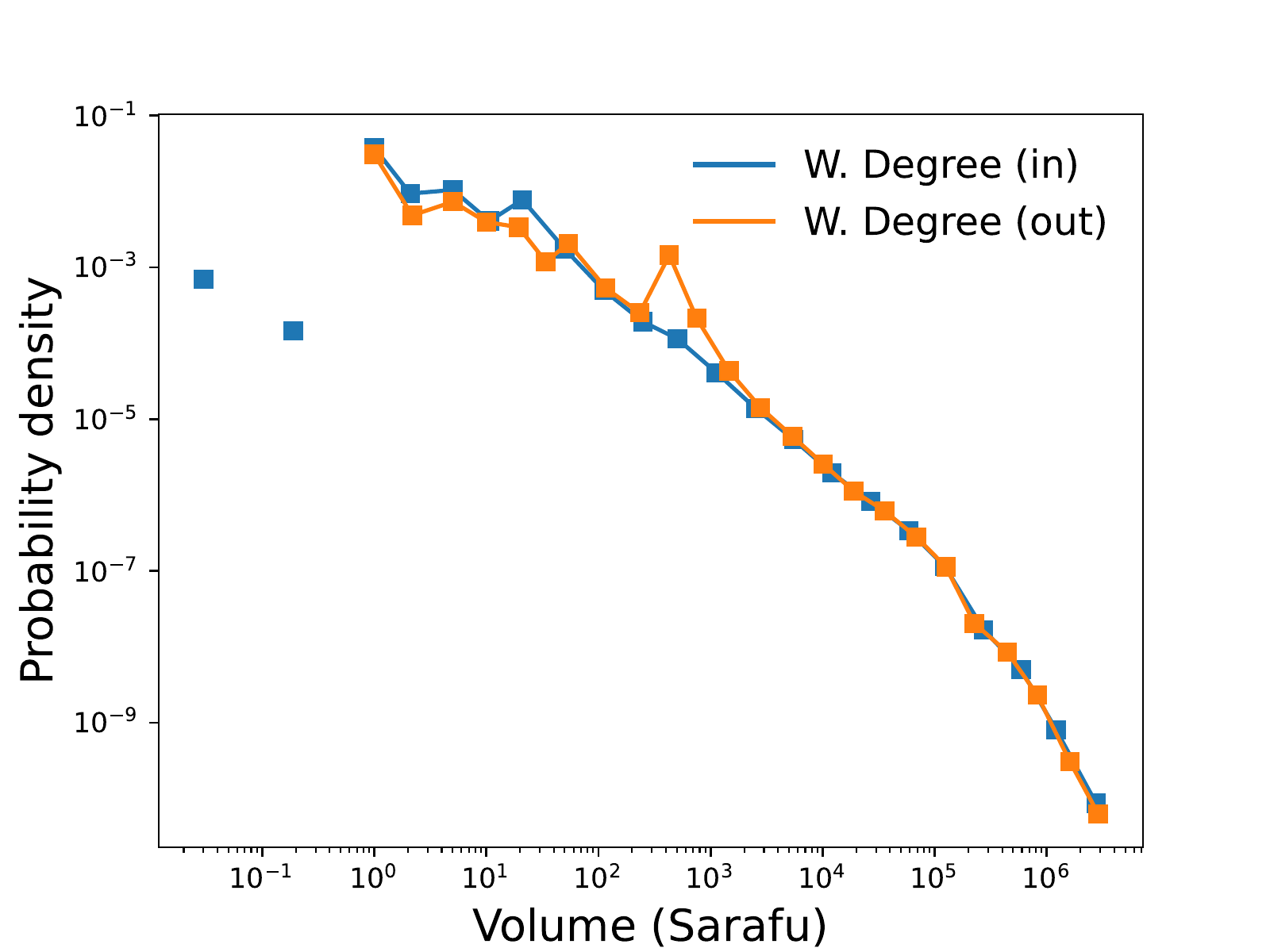}
  \caption{Distribution of degree (left) and weighted degree (right) for the Sarafu flow network. Probability densities are scaled such that nodes with a value of zero shrink the distribution total, as zero cannot be plotted on a logarithmic scale.}
  \label{fig:degrees}
\end{figure}

As a first step towards understanding prominent Sarafu users, we consider distributions of relevant account statistics. Figure~\ref{fig:degrees} (left) shows the empirical distributions of node degree, binned on a logarithmic scale. Node degree corresponds to an account's number of unique transaction partners. We note that values are highly heterogeneous across accounts; the distribution is right-skewed and there is a ``heavy'' or ``thick'' tail indicating that a small share of accounts has orders of magnitude more unique transaction partners than do most accounts. Indeed, the in- and out- high-degree tails both fit a Pareto distribution with power-law exponent of 2.9. For details we refer to the \nameref{sec:methods:analysis:tail-estimation} section of \nameref{sec:methods}.

The Sarafu flow network has hubs, in that a small share of nodes are especially prominent. Moreover, the power-law tail indicates the likely presence of so-called ``preferential attachment'' as the Sarafu network grew\cite{barabasi_emergence_1999,barabasi_network_2016,lynn_emergent_2022}. Under such dynamics, accounts with many unique transaction partners are proportionately more likely to attract or create additional connections. Because degree is disassortative (see the \nameref{sec:results:structure:mixing} section), we also know that high-degree nodes are dispersed across the network and likely to be what we call ``local hubs'' prominent among predominantly low-degree neighbors.

We also see that a relatively small number of account holders spend orders of magnitude more Sarafu than do the bulk of the users. Figure~\ref{fig:degrees} (right) shows binned empirical distributions of weighted degree, which corresponds to total transaction volumes into and out of accounts. These values are spread over a wide range and also exhibit a heavy tail. Reflecting the underlying accounting consistency present in networks of monetary flow, weighted in- and out- degree are exceptionally highly correlated. Accounts must receive large amounts of Sarafu in order to spend large amounts of Sarafu. 

\begin{figure}[!b]
  \centering
  \includegraphics[width=0.45\textwidth]{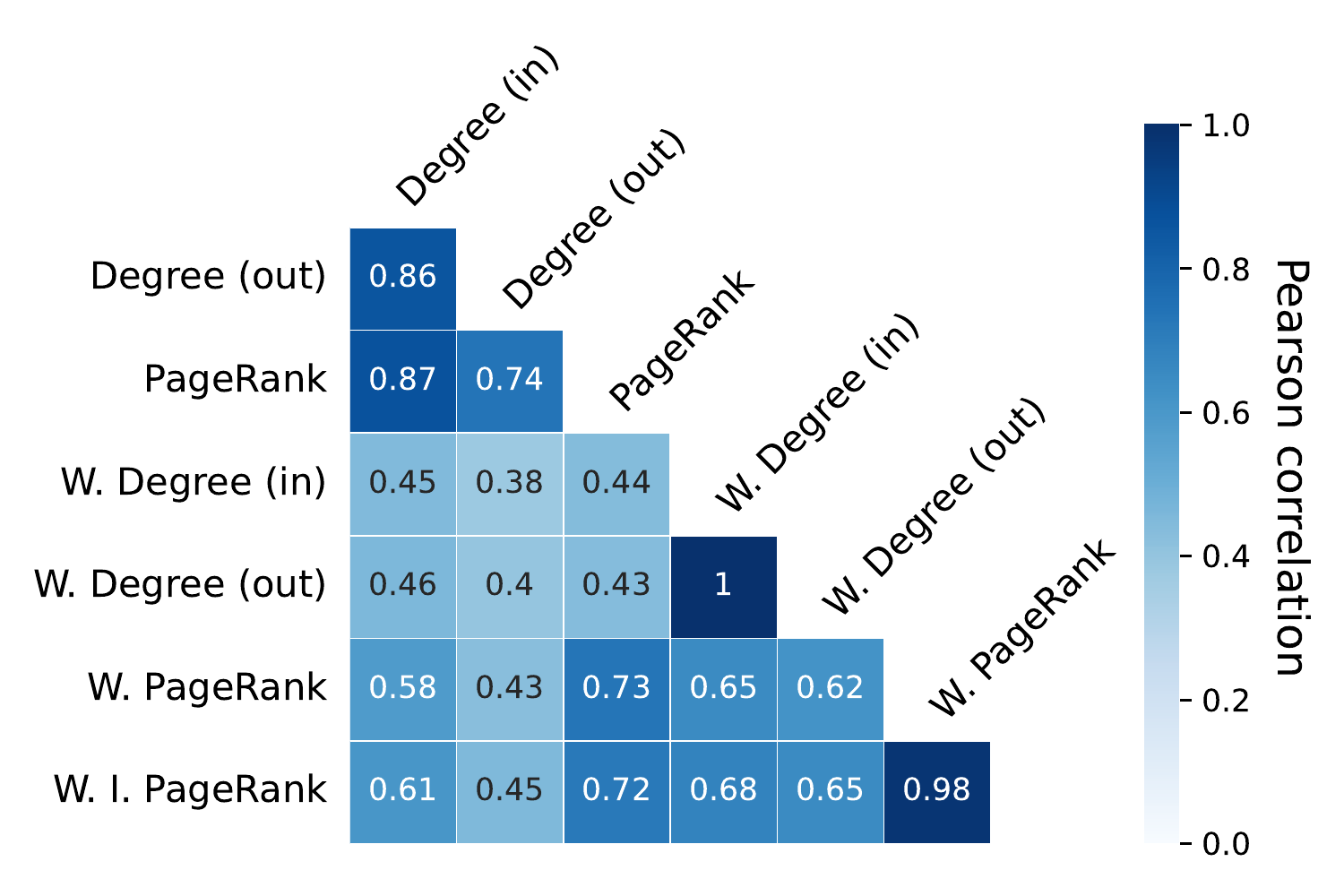}
  \caption{Pearson correlation between values for degree, weighted degree, and centrality metrics.}
  \label{fig:correlations}
\end{figure}

More generally, degree and weighted degree are not interchangeable, capturing different notions of prominence in the system. Figure~\ref{fig:correlations} shows the Pearson correlation between node degree, weighted degree, and two especially relevant centrality metrics. The unweighted PageRank algorithm produces a non-zero value for each node that is correlated with both the in- and out- degree; this makes it a practical measure for downstream tasks involving the unweighted network. Unweighted PageRank is closely related to the indegree\cite{litvak_in-degree_2007,fortunato_approximating_2008}. More interesting is the weighted PageRank, which captures something distinct from the in- or out- degree, the weighted in- or out- degree, and unweighted PageRank. Noteworthy is that values of weighted PageRank are interpretable as the share of system funds that an account would eventually control, if the observed dynamics were to continue. Further details, including an empirical calibration to account balances, are presented in the \nameref{sec:methods:analysis:centrality} section in \nameref{sec:methods}.

\subsubsection{Characterizing prominent users} \label{sec:results:hubs:users}

\begin{figure}[!t]
  \centering
  \includegraphics[width=0.8\textwidth]{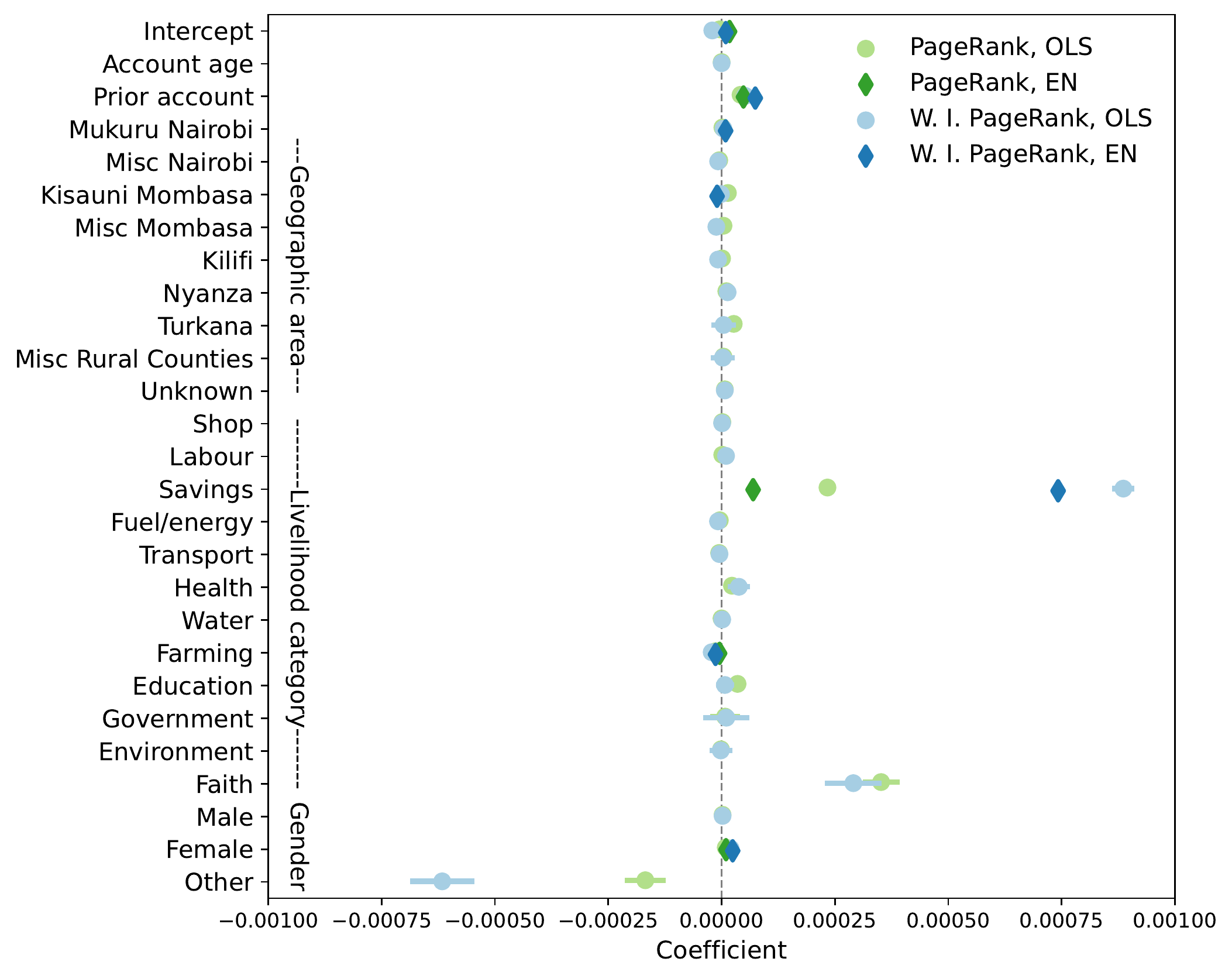}
  \caption{Regression coefficients for linear models fitting account features to centrality measures, using Ordinary least squares (OLS) and Elastic Net (EN). For the three categorical predictors, the reference categories are accounts that report a location in \textit{Kinango Kwale}, report selling \textit{food}, and do not report a gender.}
  \label{fig:regression}
\end{figure}

To characterize prominent users of the Sarafu system, we ask what features are especially consistent among accounts with high network centrality. Figure~\ref{fig:regression} illustrates the regression coefficients on account properties when PageRank and weighted, inflow-adjusted PageRank are used as outcome variables. Ordinary least squares (OLS) provides an estimated statistical contribution for each account feature, while Elastic Net (EN) incorporates regularization to highlight only those features most consistently associated with centrality. For details, we refer to the \nameref{sec:methods:analysis:regression} section in \nameref{sec:methods}.

PageRank and weighted inflow-adjusted PageRank capture distinct aspects of node importance, but are positively and negatively associated with many of the same account features. Most strongly and consistently associated with high network centrality are accounts held by \textit{savings} groups. Indeed, community-based savings and investment groups are a key feature of local economies in Kenya and of the Sarafu system (as noted in the \nameref{sec:methods:data} section in \nameref{sec:methods}). The size of this category is quite small, however, containing only 264 accounts. The number of \textit{faith} leaders is even smaller, and some appear to play an especially prominent role in the local circulation supported by this community currency. 

The other regression coefficients in Figure~\ref{fig:regression} reveal additional nuances among some of the largest categories of users. Accounts that were created \textit{prior} to the consolidation of Sarafu, which occurred as the data collection period began, are consistently associated with high network centrality; early adopters show a higher tendency to be prominent users. This is consistent with the presence of preferential attachment in this system, as noted in the previous section. We also find that account holders reporting their gender as \textit{female} are associated with higher centrality in the Sarafu flow network---this prominence of women is remarkable. In fact, it conforms to qualitative accounts from field studies in Kinango, Kwale that report strong participation of women, and women's leadership, within community-based savings and investment groups that use Sarafu\cite{avanzo_relational_2019}. This has also been noted about savings groups in Kenya, more generally\cite{rasulova_impact_2017}. With respect to geography, recall that \textit{Mukuru Nairobi} and \textit{Kisauni Mombasa} refer to the site of targeted introductions in spring 2020 and early 2021, respectively. Timing appears to have made a substantial difference: the second intervention did not spur large transaction volumes, while those reporting a location within \textit{Mukuru Nairobi} have higher network centrality (on average) than users in \textit{Kinango Kwale} (the reference category). Perhaps most interestingly, \textit{farming} is associated with lower centrality than other reported economic activities. Non-farming activities (e.g. selling \textit{food}, running a \textit{shop}, or providing \textit{labour}) appear to be ``central'' to local economies even in areas of rural Kenya, such as \textit{Kinango Kwale}, that rely heavily on subsistence agriculture.

\section{Discussion} \label{sec:discussion}

Our study of the Sarafu system complements and corroborates a growing body of work informing policy on alternative interventions in marginalized areas~\cite{ruddick_complementary_2015,mauldin_local_2015,fuders_smarter_2016,gomez_monetary_2018,ussher_complementary_2021}. In particular, our findings shed new light on the conditions under which community currencies might form part of successful humanitarian or development interventions. The circulation of Sarafu remained modular and local even as the system saw dramatic growth in its user base and accommodated large spikes in transaction volumes. In areas where local economic activities are already diverse, the rapid deployment (or scaling up) of community currency operations may be able to support the aims of a broader humanitarian response. Over longer periods of time, and in more peripheral areas, community currencies support economic development to the extent that they encourage diverse productive activities and strengthen short, local supply chains that keep money within a community~\cite{ussher_complementary_2021}. Here it is useful to note that community-based financial institutions, and, in some cases, faith leaders, played a key role in the Sarafu system. Future initiatives would likely benefit from finding ways to better assist in the discovery or creation of short cycles in local economies. Practically speaking, it may be possible to identify ``missing links'' in local economic or financial networks such that policymakers and organizers might intervene to close cycles by brokering among local businesses~\cite{fleischman_balancing_2020,fleischman_liquidity-saving_2020}.

More broadly, this work has demonstrated that a network approach can unveil meaningful patterns and extract relevant insights about the circulation of money from digital transaction records. Our contextually relevant findings demonstrate the explanatory power of representing circulation as a network of monetary flow. When applied to such networks, walk- and flow- based network analysis methods (e.g., PageRank and Infomap) produce outputs that are readily interpretable in terms of circulation and account balances. Notably, these methods rely on scalable algorithms meaning that our approach can be applied to study sizeable currency systems where transaction data is recorded in digital form. This includes other community currencies~\cite{muralt_woergl_1934,kichiji_network_2008,frankova_transaction_2014,iosifidis_cyclic_2018} as well as major global cryptocurrencies~\cite{kondor_rich_2014,elbahrawy_evolutionary_2017,nadini_emergence_2022,campajola_microvelocity_2022}. Recent methodological advances~\cite{mattsson_trajectories_2021,kawamoto_single-trajectory_2022} promise to extend applicability also to payment systems that are not themselves full currency systems, such as mobile money systems~\cite{blumenstock_airtime_2016,economides_mobile_2017,mattsson_trajectories_2021}, large value payment systems~\cite{soramaki_topology_2007,iori_network_2008,kyriakopoulos_network_2009,bech_illiquidity_2012,barucca_organization_2018,rubio_classifying_2020,bianchi_longitudinal_2020}, major banks~\cite{zanin_topology_2016,rendon_de_la_torre_topologic_2016,carvalho_tracking_2021,ialongo_reconstructing_2021}, and, in an exciting development, centralized national payment infrastructures~\cite{triepels_detection_2018,sabetti_shallow_2021,arevalo_identifying_2022} or central bank digital currencies~\cite{bank_of_canada_central_2021}.

This study considers the aggregated circulation of Sarafu over the full observation period, and applies especially suitable network analysis techniques using openly available tools. While this already presents interesting findings, networks of monetary flow are static representations and potentially interesting temporal or sequential information is lost\cite{saramaki_exploring_2015,mattsson_trajectories_2021,larock_sequential_2022,mattsson_measuring_2022,campajola_microvelocity_2022}. Several changes to the system occurred within this period and could be studied in future work\cite{mattsson_sarafu_2022}. Transaction processes are compelling examples of real-world walk processes~\cite{mattsson_trajectories_2021}, and we hope that the interpretability of walk-based network analysis methods as applied to our static representation of Sarafu can help motivate the development of equally interpretable methods for dynamic representations. It may be possible, for instance, to produce measures of the fidelity between the temporal network that fully describes a transaction process and its static representation, as has been done for interaction networks in the context of spreading processes\cite{lentz_unfolding_2013}. Indeed, reproducing observed trajectories (from, for example, a real-world walk process) using a static network representation is already the logic underlying multi-order models of complex systems~\cite{xu_representing_2016,lambiotte_networks_2019,mattsson_trajectories_2021}. Future methodological advances would also be welcome in the analysis of network cycles and of log-normal distributions with Pareto tails.

Studies on additional systems and further methodological development could bring us towards a future where entire economies might be studied as (interconnected, dynamic) networks of monetary flow. Notably, degree disassortativity has been found across many economic networks~\cite{fujiwara_large-scale_2010,kondor_rich_2014,mattsson_functional_2021,campajola_microvelocity_2022} and short cycles are over-represented also in at least one other complementary currency system~\cite{iosifidis_cyclic_2018} despite considerable contextual differences. This is some indication that there may be important network-structural regularities underlying the circulation of money, which deserve to be further explored across currency systems large and small.

\section{Methods} \label{sec:methods}

The \nameref{sec:methods:data} section provides a detailed description of the portion of the raw Sarafu data used in constructing our timeseries and our network of monetary flow, plus three peculiarities of the Sarafu currency system that are of relevance to the implementation or interpretation of our analyses. Network analysis methods are used to quantitatively analyze the Sarafu flow network. The \nameref{sec:methods:analysis:circ} section articulates how the map equation framework captures and quantifies the circulation of money given a network of monetary flow. The \nameref{sec:methods:analysis:centrality} section describes walk-based measures of network centrality for characterizing prominent users. The \nameref{sec:methods:analysis:cycles} and \nameref{sec:methods:analysis:mix} sections introduce relative cycle counts and assortativity as tools to analyse the structure of the underlying, unweighted network. 

\subsection{Data preparation} 
\label{sec:methods:data}

The Sarafu CIC data\cite{ruddick_sarafu_2021} includes a transaction dataset and an account dataset collected from January 25th, 2020 to June 15th 2021. The raw form of this data has previously been described in detail\cite{mattsson_sarafu_2022}. The transaction records are labeled with a transaction type, and we consider the \textsc{standard} transactions. Figure~\ref{fig:sarafu} shows the total volume of such transactions for each complete month. Note that the value of one Sarafu was understood by users to be about that of a Kenyan Shilling, though actual exchange was facilitated only in very limited instances. The Sarafu flow network is constructed from the \textsc{standard} transactions that occurred within the Sarafu system over the observation period. Basic network statistics are shown in Table~\ref{tab:network}. As noted in the main text, the \emph{nodes} are registered accounts, for which the account dataset includes relevant account features (detailed below). An \emph{edge} from one account to another indicates that at least one \textsc{standard} transaction occurred across that link. The \emph{edge weight} corresponds to the total sum of all \textsc{standard} transaction amounts across that link. Then, system-run accounts are filtered out from the Sarafu flow network. Regular accounts who neither sent nor received even a single \textsc{standard} transaction from another regular account are isolates, which we also exclude from the network. Note that the giant connected component (GCC) encompasses nearly all the nodes, meaning that the majority of users are indirectly connected through their transactions.

\begin{table}[htb]
    \centering
    \begin{tabular}{l|rrrr}
         & Nodes & Edges & Transactions & Volume (Sarafu) \\ 
        \textsc{standard} transactions & 40,767 & 146,615 & 422,721 & 297.0 million \\
        Sarafu flow network & 40,657 & 145,661 & 421,329 & 293.7 million \\
        GCC & 38,653 & 143,724 & 418,675 & 293.4 million
   \end{tabular}
    \caption{\label{tab:network}Basic statistics for the network of aggregated \textsc{standard} transactions, the Sarafu flow network, and its giant connected component (GCC).}
\end{table}

\medskip
\noindent
\textbf{Account features}.
The account dataset includes the registration date and reported gender of the account holder as well as categorical labels derived from reported information on home location and livelihood. Mattsson, Criscione, \& Ruddick\cite{mattsson_sarafu_2022} provide a descriptive overview of each account feature. Notably, each geographic area is a combination of user-reported localities that could be quite precise. Ussher et al.\cite{ussher_complementary_2021} present an overview of the user-reported work activities that make up the livelihood categories. System-run accounts are those labeled with \textit{system} in place of the node attribute indicating the user's livelihood, or assigned a formal role as an \textsc{admin} or \textsc{vendor} account. 

\medskip
\noindent
\textbf{Savings \& investment groups}. 
Community-based savings and investment groups are common in Kenya~\cite{biggart_banking_2001,central_bank_of_kenya_inclusive_2019} and a feature of many communities that use Sarafu, specifically~\cite{marion_voucher_2018,ussher_complementary_2021}. Several hundreds of so-called ``chamas'' are present in the data, many with the label \textit{savings} in place of the node attribute indicating the user's livelihood. For a time, Sarafu operator Grassroots Economics also had a program whereby field staff would verify the operation of community-based groups and provide additional support to verified ``chamas''\cite{mattsson_sarafu_2022}. Verified groups were conduits for development initiatives and humanitarian aid on several occasions, some of which involved payments made to system-run accounts, in Sarafu, in exchange for donated food, items, or Kenyan Shillings.

\medskip
\noindent
\textbf{Targeted introductions}. 
There were two so-called targeted introductions during the observation period, conducted by the Kenyan Red Cross in collaboration with Grassroots Economics~\cite{mattsson_sarafu_2022}. These consisted of outreach efforts and training programs in specific areas. The Mukuru kwa Njenga slum in Nairobi was the site of the first; educational and outreach programs began in April 2020. Soon thereafter, this intervention was scaled up in response to the COVID-19 pandemic and related economic disruptions. Again, community currencies tend to gain in popularity during times of economic and financial crisis~\cite{stodder_complementary_2009,stodder_macro-stability_2016,zeller_economic_2020}. A second Red Cross intervention began in in Kisauni, Mombasa in early 2021. This resulted in a wave of account creations~\cite{mattsson_sarafu_2022} and rising activity by accounts with location \textit{Kisauni Mombasa}~\cite{ussher_complementary_2021}. However, as we can see in Figure~\ref{fig:sarafu}, overall transaction volumes did not rise as dramatically during this targeted introduction as they did during the first.

\medskip
\noindent
\textbf{Currency creation}.
The digital payment system, as a whole, saw inflows when new units of Sarafu were created. For instance, newly-created accounts would receive an initial disbursement of 400 Sarafu, later reduced to 50 Sarafu. New users could receive an additional sum if and when they verified their account information with staff at Grassroots Economics. Existing users could also receive newly created funds, such as in reward for transaction activity and as bonus for referring others. These and other non-\textsc{standard} inflows are summarized as an aggregated value in the account dataset. We refer to prior work for a full account of currency management and system administration over the data collection period\cite{mattsson_sarafu_2022}. 

\subsection{Circulation} \label{sec:methods:analysis:circ}


To study circulation we turn to the map equation framework~\cite{rosvall_maps_2008} and the associated Infomap algorithm~\cite{ding_community_2014}. This is an approach based on computations involving a walk process over a given network, which is relevant in that financial transactions describe a real-world walk process\cite{mattsson_trajectories_2021}. The edge weights of the Sarafu flow network already reflect flows and so simulating a walk process is not necessary. Infomap takes a weighted, directed network as input and outputs a hierarchical mapping of nodes grouped into discovered \emph{modules}. This grouping is done via computational optimization. Specifically, the map equation defines a notion of entropy whose value is higher the more of the flow over the given network occurs between rather than within modules. The Infomap algorithm exploits meso-scale network structure to minimize that value, grouping together nodes with much flow among themselves (and little outside). These are discovered sub-populations among whom a ``random walker'' would tend to stay for relatively long. We refer to top-level modules, sub-modules and sub-sub-modules at respectively the first, second and third level of the discovered hierarchy. The composition of these sub-populations can then be understood by means of an approach where we quantify their heterogeneity along dimensions of geography, livelihood, and gender, i.e., the node attributes. Implementation details for running Infomap and analyzing the resulting module mapping are included in Supplementary File 2. 

\subsection{Network cycles} \label{sec:methods:analysis:cycles}

To describe the network connectivity patterns underlying the circulation of Sarafu, we consider cycles. A \textit{cycle} is a simple path starting and ending at the same node. In the context of complementary currencies, cycles ensure the flow of liquidity throughout the system. For cycles to occur, users must be willing to both spend and earn in complementary currency. Following this observation, we analyze cyclic structures in the Sarafu network using the $z$-score of the empirical cycle count as compared to a null model. We consider two common null models: Erdős-Rényi (ER) networks and random degree-preserving (RD) networks. ER networks have the same number of nodes and edges as the empirical network, but are wired randomly. Our RD networks preserve the indegree and outdegree sequences of the empirical network in expectation, but are otherwise random. Specifically, we sample from a binary, directed Exponential Random Graph Model\cite{vallarano_fast_2021}. The $z$-score is defined in relation to the mean and standard deviation of the number of cycles of length $k$ detected in 100 realizations of the relevant null model.

The cycle $z$-score is computed separately for each Infomap sub-module composed of 100 or more accounts (see: \nameref{sec:methods:analysis:circ}). Directed cycles are detected and counted using an existing approach~\cite{butts_cycle_2006}. This is done for each empirical sub-network, and for the ER and RD graphs generated from the empirical sub-network. An implementation is provided in Supplementary Files 4 and 5. 

\subsection{Network mixing patterns} \label{sec:methods:analysis:mix}

To characterize the mixing patterns underlying the network structure of the Sarafu community currency, we consider degree and attribute assortativity~\cite{newman_mixing_2003,foster_edge_2010}. Values are computed separately for each sub-network delineated by the Infomap sub-modules composed of 100 or more accounts (see: \nameref{sec:methods:analysis:circ}). The categorical attribute assorativity is calculated along dimensions of livelihood category and reported gender, using the undirected version of the networks. These measures compare the number of links between accounts with the same livelihood or gender to that which would be expected at random, and can range from -1 to 1. A value of 0 corresponds to random expectation; a value of 1 corresponds to a network where transactions only occurred between accounts with the same attribute value. When there is no variation within a sub-population, the sub-network is given an assortativity value of 1. Similarly, we calculate the numerical attribute assorativity to quantify mixing patterns with respect to registration date, in-degree, and out-degree. Implementation details are reported in Supplementary File 2.

\subsection{Network centrality}
\label{sec:methods:analysis:centrality}

To characterize prominent users of Sarafu, we employ network centrality measures on the Sarafu flow network. Purely structural node-based metrics such as degree and weighted degree correspond to straightforward statistics about accounts. We also use walk-based methods for node centrality as these are especially interpretable with respect to monetary flow; the well-known PageRank algorithm is flexible and computationally tractable. These centrality measures are computed for our network, and then interpreted in the context of node attributes of the account-holders using linear regression. This lets us characterize prominent users without highlighting individual account holders, which is neither our goal nor desirable for privacy reasons (see the \nameref{sec:methods:data_availability} section). Below, we briefly discuss each employed measure.

\textbf{Indegree and outdegree} in the Sarafu flow network correspond to an account's number of unique incoming and outgoing transaction partners, respectively, over the observation period. It is possible for nodes to have zero indegree \emph{or} outdegree, but accounts with neither incoming nor outgoing transaction partners would be isolates and these are filtered out. 

\textbf{Weighted indegree and weighted outdegree} in the Sarafu flow network correspond to total transaction volumes into and out of accounts over the observation period. Note that these are mechanistically related in that money must be obtained before it can be spent according to the accounting consistency enforced within payment systems.

\textbf{PageRank} is an algorithm that produces a walk-based metric for node centrality given a directed network~\cite{page_pagerank_1999,frankova_transaction_2014}. The obtained centrality values approximate the probability of finding a random walker at a given node at any given moment. More specifically, PageRank computes the stationary probability of a random walk process with restarts on a given network. A single parameter $\alpha$ is used to control the propensity for the simulated walkers to restart. An $\alpha$ value of $0.85$ is the long-established default, meaning that $15\%$ of times a random walker will restart rather than follow an out-link from the node where it currently resides. By default, restarts are uniformly random across the nodes. However, it is also possible to specify the probability of restarting at any particular node using a so-called personalization vector.

\textbf{Weighted PageRank} is an analogous centrality metric for weighted, directed networks. Over a weighted network, the random walkers choose among available out-links in proportion to their edge weights. Weighted edges are flows of money, in our case, and so the stationary probability of a random walker would then corresponds to the share of the total balance held by each account at equilibrium. This makes the the Weighted PageRank algorithm especially applicable in the context of a currency system context, since it means that the values are interpretable as the share of system funds that an account would eventually control if observed dynamics continued. Within this intuition, \textbf{Weighted Inflow-adjusted PageRank} employs a personalization vector to better capture idiosyncratic patterns of currency creation; real-world currency systems may be poorly represented by the default assumption of uniformly random restarts. Recall from the \nameref{sec:methods:data} section that Sarafu users could receive disbursements and rewards in addition to inflows from regular transactions. We use the aggregated values of non-\textsc{standard} inflows, available in the account dataset, to set the PageRank personalization vector. The simulated random walk process is then constrained to reproduce the observed pattern of currency creation, on average.

\subsubsection{Empirical calibration}
\label{sec:methods:analysis:centrality:calibration}

Running the Weighted PageRank algorithm requires specifying the aforementioned parameter $\alpha$. We would like to understand whether this parameter affects the suitability of these values as a centrality measure for networks of monetary flow. Since Weighted PageRank extrapolates the observed patterns of circulation towards a future where an equilibrium is reached, the output values can be understood as a prediction for hypothetical future account balances (as a fraction of the total balance). While we cannot expect such strong modeling assumptions to produce especially accurate estimates, it is nonetheless instructive to compare to empirical account balances. We can determine whether this centrality metric is sensitive to $\alpha$ and whether modeling non-random currency creation, via the PageRank personalization vector, matters for this particular real-world system. 

\begin{figure}[!t]
    \centering
  \includegraphics[width=0.45\textwidth]{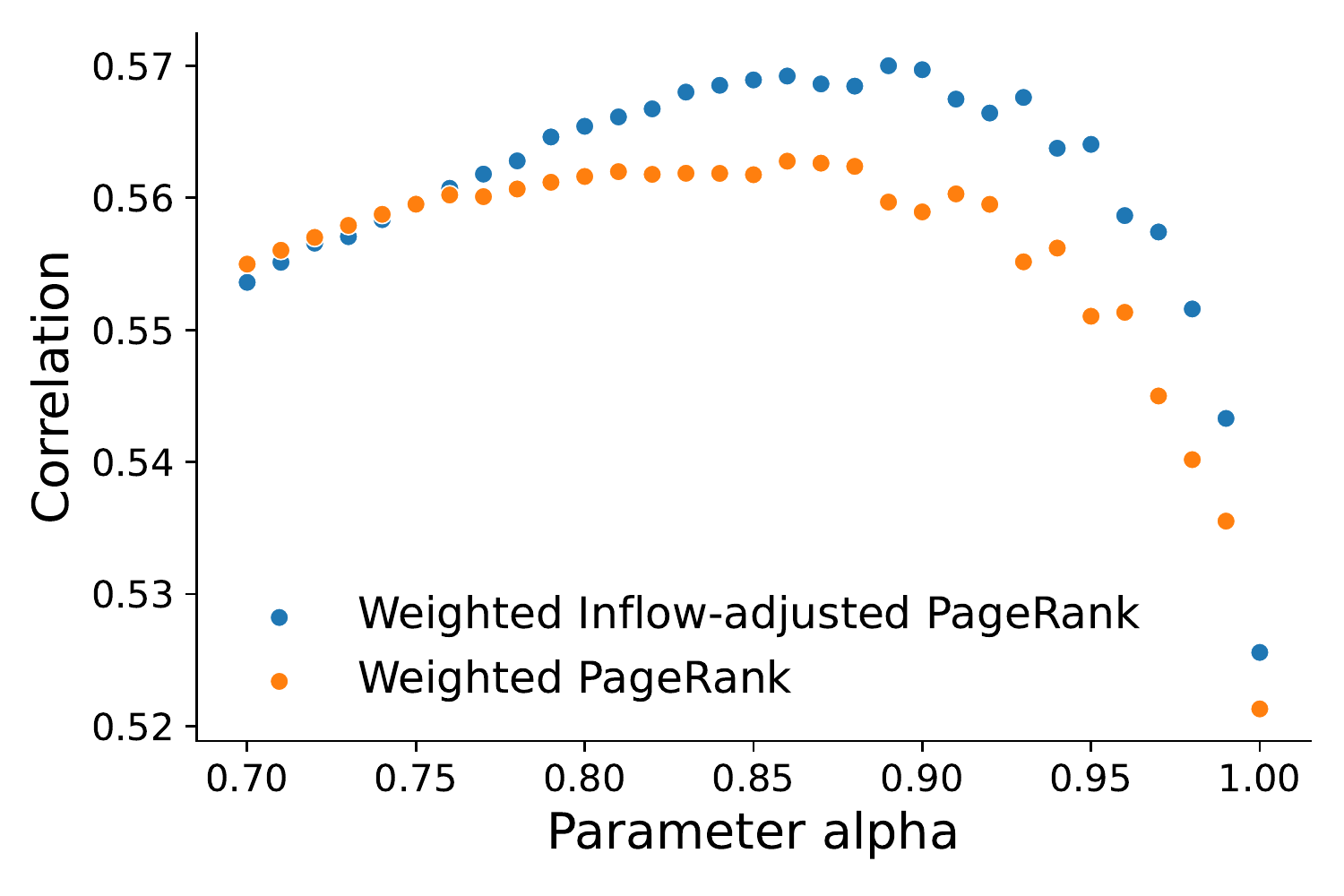}
  \caption{Pearson correlation of Weighted and Weighted Inflow-adjusted PageRank with final account balances.}
  \label{fig:alphas}
\end{figure}
 
We consider the correlation between our centrality metrics computed on the Sarafu flow network with Sarafu balances observed at the time of data collection on June 15th, 2021. Figure~\ref{fig:alphas} plots the correlation between these final balances and the values given by the Weighted PageRank algorithm, with and without adjusting the simulated walk process to account for currency creation. The resulting correlations are at most $R = 0.57$ and $R = 0.56$, respectively. Taking the perspective that Weighted PageRank estimates hypothetical future account balances, it is encouraging to note that these values correlate more closely with final balances than do the in- or out- degree ($R = 0.28$, $R = 0.21$), and the weighted in- or out- degree ($R = 0.52$, $R = 0.47$). Moreover, both versions of Weighted PageRank produce values that are consistently correlated with final balances over a wide range of parameter values that includes the long-established default ($\alpha = 0.85$); our centrality metrics are not overly sensitive to the propensity for restarts. We conclude that Weighted PageRank, especially Weighted Inflow-adjusted PageRank, is a highly suitable centrality metric for downstream analyses of networks of monetary flow.

\subsection{High-degree tail estimation} \label{sec:methods:analysis:tail-estimation}

To estimate the high-degree tail exponent of the degree distributions, we apply a recent approach based in extreme value theory\cite{voitalov_scale-free_2019}. This approach posits that heavy-tailed degree distributions might be usefully described as so-called ``regularly varying'' distributions, following a power-law at high degrees even if they take on an \textit{arbitrary functional form} at lower degrees. Specifically, regularly varying distributions have a probability density function (PDF) of the form $P(k)=\ell(k) k^\gamma$, where $\ell(k)$ is a slowly varying function. The PDF tail exponent $\gamma$, the complementary cumulative distribution function (CCDF) tail exponent $\alpha=\gamma-1$, and the so-called tail index $\xi = \frac{1}{\alpha}$ can be estimated using established techniques. Regularly varying distributions have a tail index greater than zero, that is, $\frac{1}{\gamma-1}=\frac{1}{\alpha}=\xi>0$.

\begin{figure}[!t]
  \includegraphics[width=0.32\textwidth]{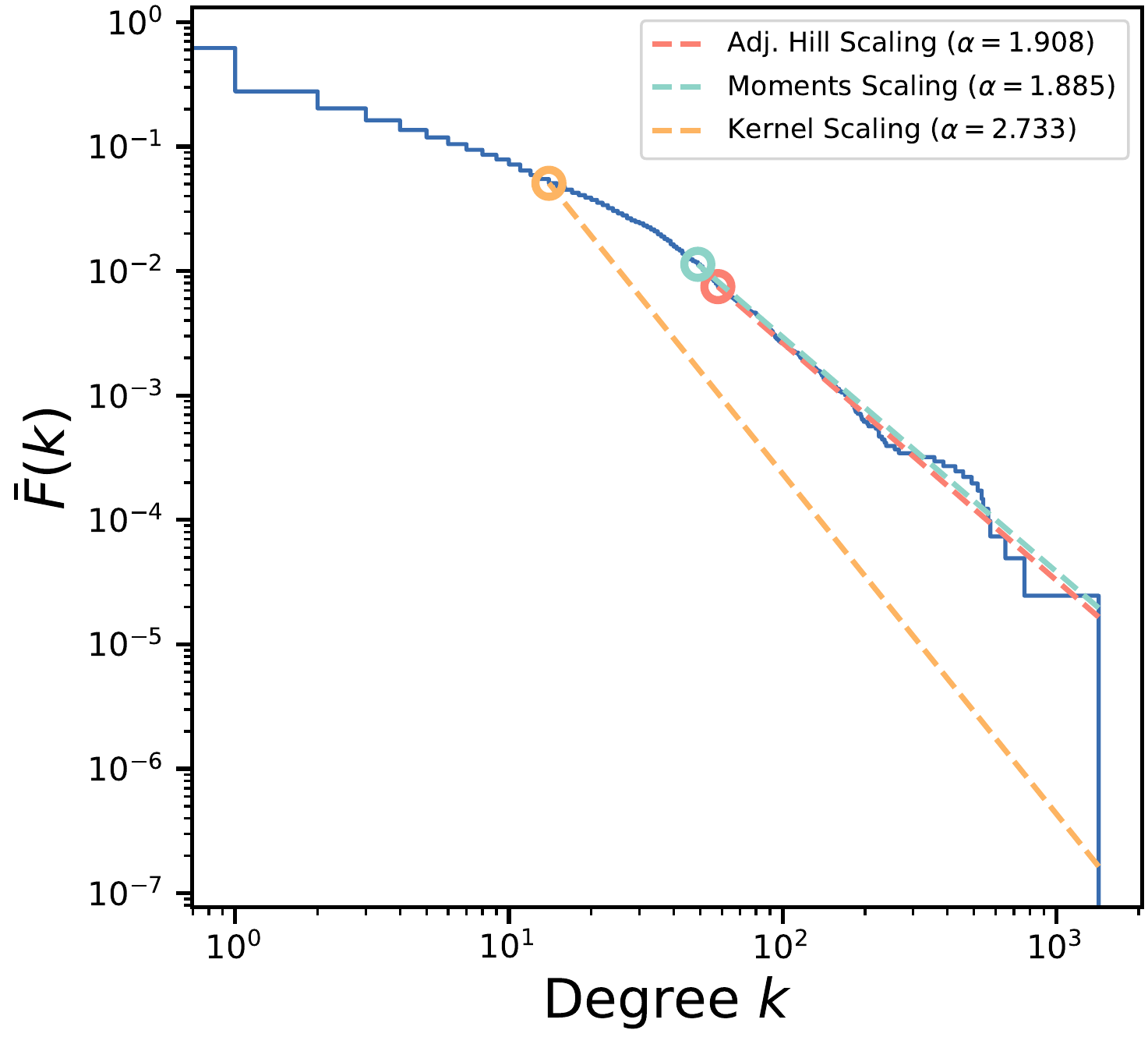}
  \includegraphics[width=0.32\textwidth]{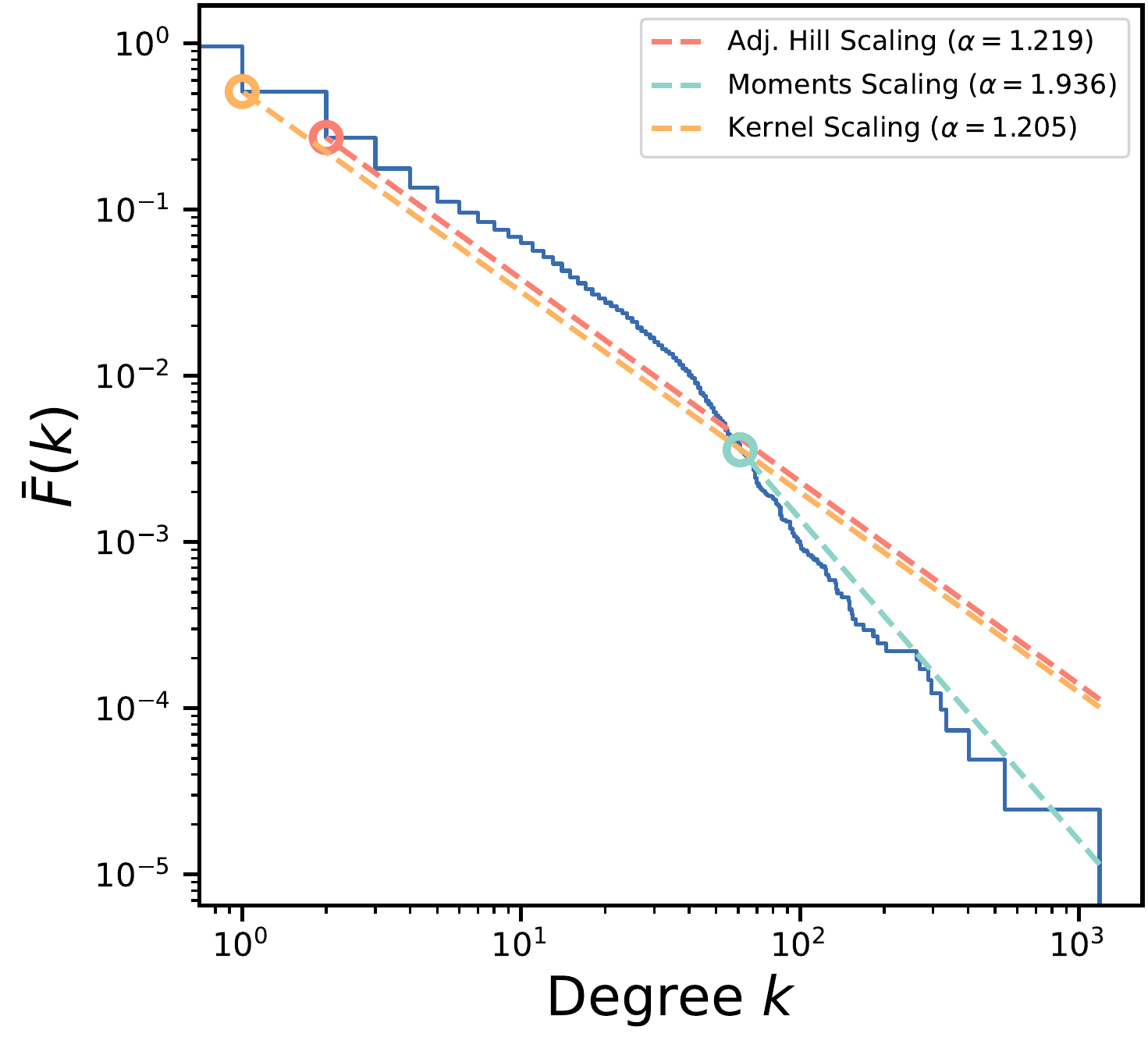}
  \includegraphics[width=0.32\textwidth]{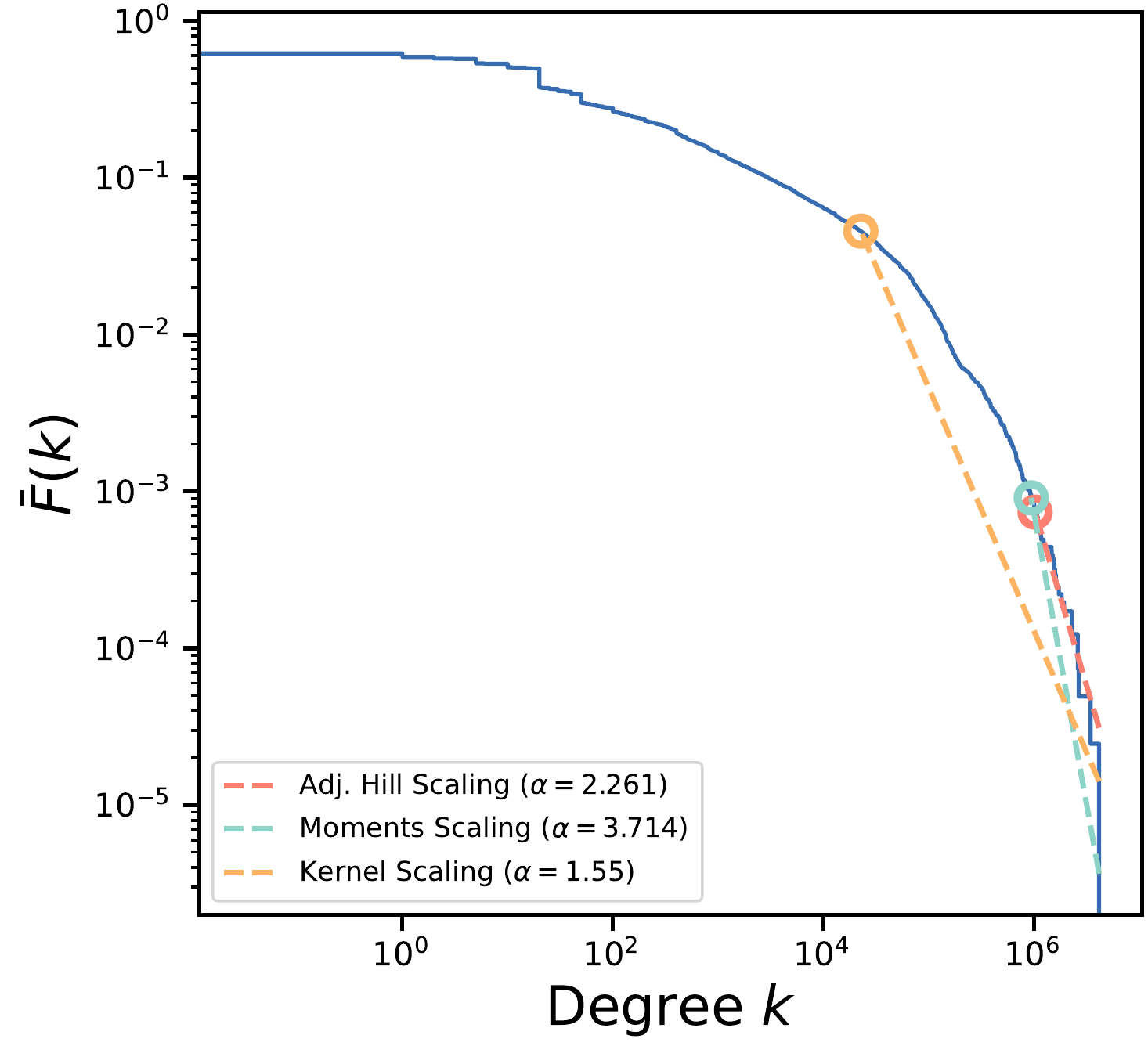}
  \caption{Complementary cumulative distribution functions (CCDF) and fitted tail exponents for the indegree, outdegree, and weighted indegree of the Sarafu flow network. Weighted outdegree is not shown.}
  \label{fig:tail-estimation}
\end{figure}

The high-degree tail exponent can be clearly estimated for our indegree and outdegree sequences, using the automated and openly available fitting procedure\cite{voitalov_tail_2022} provided alongside Voitalov et. al. (2019) \cite{voitalov_scale-free_2019}. Figure~\ref{fig:tail-estimation} reproduces the output of the fitting procedure, wherein the empirical CCDF is plotted against tail fits by the three available estimators. The Moments estimator appears well-behaved for both indegree (left) and outdegree (center); the full diagnostic plots are included in Supplementary File 6. Notably, the other estimators also show local minima with respect to the fitting criteria at or near the tail-cutoff selected by the Moments estimator. We conclude that the best estimate of $\gamma$ is around 2.9 for the in- and out- high-degree tails. Lowering the tail cutoff to include the bulk of the outdegree distribution produces a lower estimate for $\gamma$, at around 2.2.

The high-degree tail exponents for the weighted degree distributions, however, cannot be estimated in this straightforward way. Figure~\ref{fig:tail-estimation} (right) shows the empirical CCDF of the weighted indegree and the fitted high-degree tail of the CCDF as estimated by the three available estimators. Recall that weighted in- and out- degree are closely correlated due to the underlying accounting consistency of the system (see the \nameref{sec:results:hubs:centrality} section). In both cases, the fitted exponent varies smoothly with the fraction of nodes included in the high-degree tail. This may be due to the inherent difficulty of empirically distinguishing Pareto from log-normal distributions\cite{mitzenmacher_brief_2004,broido_scale-free_2019}. Indeed, the weighted in-degree of the Sarafu flow network corresponds to some notion of ``income'' within this system and, at the national scale, such distributions are often described as log-normal with a Pareto tail\cite{souma_universal_2001,reed_double_2004,clementi_paretos_2005,battistin_why_2009}. Additional approaches, such as finite size scaling analysis~\cite{serafino_true_2021}, would be needed to determine the functional form and estimate the high-degree tail exponents for the weighted degree distributions.

\subsection{Linear regression} \label{sec:methods:analysis:regression}

To assess what recorded features of the account holders associate with higher prominence, as measured by network centrality, we use linear regression. Ordinary least squares (OLS) is used to fit a linear model to an outcome, in our case a network centrality measure, providing an estimated contribution for each input feature~\cite{montgomery_introduction_2012}. Regularization is a fitting technique that introduces a penalty term to the optimization limiting the number of regressors and/or their magnitudes\cite{friedman_regularization_2010}. So-called Elastic Net (EN) regularization, as we use it, penalizes the number of regressors and their magnitude equally. The penalty weight is selected using five-fold cross validation, just before the point where additional features begin entering the model without qualitatively improving the statistical fit. Further implementation details are noted in Supplementary File 3, alongside the code that replicates the analysis. 

\subsection{Data availability} 
\label{sec:methods:data_availability}
The dataset analyzed in this study~\cite{ruddick_sarafu_2021} is available via the UK Data Service (UKDS) under their End User License, which stipulates suitable data-privacy protections. The dataset is available for download from the UKDS ReShare repository (\url{https://reshare.ukdataservice.ac.uk/855142/}) to users registered with the UKDS (\url{https://beta.ukdataservice.ac.uk/myaccount/credentials}). Further usage notes and an extensive description of the dataset are available in a complementary publication~\cite{mattsson_sarafu_2022}.

\subsection{Software availability} All software used in this study are available under an open-source licence: 
\begin{itemize}
  \setlength\itemsep{0em}

    \item \texttt{infomap v1.6.0}~\cite{edler_mapequation_2021}
    \item \texttt{networkx v2.6.3}~\cite{hagberg_exploring_2008}
    \item \texttt{netdiffuseR v1.22.3}~\cite{vega_yon_netdiffuser_2021}
    \item \texttt{sna v2.6}~\cite{butts_sna_2020}
    \item \texttt{statsmodels v0.13.2}~\cite{seabold_statsmodels_2010}
    \item \texttt{seaborn v0.11.2}~\cite{waskom_seaborn_2021}
    \item \texttt{matplotlib v3.5.2}~\cite{hunter_matplotlib_2007}
    \item \texttt{pandas v1.4.2}~\cite{reback_pandas-devpandas_2022}
    \item \texttt{NEMtropy v2.1.1}~\cite{vallarano_nemtropy_2022}
\end{itemize}

\subsection{Code availability}

The code required to construct the network and reproduce each analysis is included the Supplementary Information.
Supplementary File 1 contains code to construct the network from the raw data.
Supplementary File 2 contains code to reproduce the analysis in the \nameref{sec:results:flow} and \nameref{sec:results:structure:mixing} sections.
Supplementary File 3 contains code to reproduce the analysis in the \nameref{sec:results:hubs} section.
Supplementary File 4 contains code to reproduce the analysis in the \nameref{sec:results:structure:cycles} section.
Supplementary File 5 contains code to reproduce the figures in the \nameref{sec:results:structure:cycles} section.
Supplementary File 6 contains the full diagnostic plots referred to in the \nameref{sec:methods:analysis:tail-estimation} section and a high-resolution version of Figure~\ref{fig:network}.

\bibliography{references}

\section{Acknowledgements}

The authors thank János Kertész, Tiago P. Peixoto, Pim van der Hoorn, and William O. Ruddick for valuable feedback.

\section{Author contributions statement}

C.M. and F.W.T. developed the methods. C.M. conducted the research and drafted the initial manuscript. T.C. performed the cycle analysis and drafted the corresponding sections. All authors contributed to the final manuscript.

\section{Additional information}

The authors declare no competing interests.

\end{document}